\documentclass[a4paper,journal]{IEEEtran}

\usepackage{changepage}
\usepackage[utf8]{inputenc} 
\usepackage[T1]{fontenc}
\usepackage{url}              
\usepackage{cite}      
\usepackage{comment}
\usepackage[cmex10]{amsmath}  
\usepackage{mleftright}    
\mleftright                   

\usepackage{graphicx}      
\usepackage{booktabs}   
\usepackage{color}

\usepackage[caption=false,font=footnotesize]{subfig}

\usepackage{diagbox}
\usepackage{graphicx}
\usepackage{multirow}

\def\C {\mathbb{C}}
\def\A {\mathbb{A}}

\def\S {\mathcal{S}}
\def\D {\mathcal{D}}
\def\I {\mathcal{I}}

\def\det {\text{\rm det}\,}

\def\v {\mathbf{v}}

\usepackage{amsmath,amsfonts,bm}
\def\cleanup{\textit{clean-up}}
\def\cleanone{\textit{clean-up-\uppercase\expandafter{\romannumeral1}}}
\def\cleantwo{\textit{clean-up-\uppercase\expandafter{\romannumeral2}}}

\def\eqref#1{equation~\ref{#1}}

\def\1{\bm{1}}

\def\va{{\bm{a}}}

\def\vs{{\bm{s}}}

\def\mB{{\bm{B}}}

\def\mD{{\bm{D}}}

\def\mH{{\bm{H}}}

\def\mW{{\bm{W}}}

\DeclareMathAlphabet{\mathsfit}{\encodingdefault}{\sfdefault}{m}{sl}
\SetMathAlphabet{\mathsfit}{bold}{\encodingdefault}{\sfdefault}{bx}{n}

\def\gE{{\mathcal{E}}}

\def\gG{{\mathcal{G}}}

\def\gM{{\mathcal{M}}}

\def\gV{{\mathcal{V}}}

\newtheorem{defn}{Definition}

\newtheorem{lem}{Lemma}
\newtheorem{exe}{Example}

\begin{document}

\title{Revisiting Topological Interference Management: \\A Learning-to-Code on Graphs Perspective
} 

\author{\IEEEauthorblockN{Zhiwei Shan, \textit{Student Member, IEEE}, Xinping Yi, \textit{Member, IEEE}, Han Yu, \textit{Member, IEEE},\\ Chung-Shou Liao, \textit{Senior Member, IEEE} and Shi Jin, \textit{Fellow, IEEE}}

\thanks{This work has been presented in part at 2023 IEEE International Symposium on Information Theory (ISIT) \cite{10206636}.}
\thanks{Z.~Shan is with University of Liverpool, Liverpool L69 3GJ, United Kingdom, and also with National Tsing Hua University, Hsinchu 30013, Taiwan, ROC. Email: zshan@liverpool.ac.uk.}
\thanks{X.~Yi and S.~Jin are with National Mobile Communications Research Laboratory, Southeast University, Nanjing 210096, China. Email: \{xyi, jinshi\}@seu.edu.cn.}
\thanks{H.~Yu is with Chalmers University of Technology, Gothenburg SE41296, Sweden. Email:                  yuha@chalmers.se.}
\thanks{C.-S.~Liao is with National Tsing Hua University, Hsinchu 30013, Taiwan, ROC. Email:                      csliao@ie.nthu.edu.tw.}
}

\maketitle
\thispagestyle{empty}
\pagestyle{empty}
\begin{abstract}
The advance of topological interference management (TIM) has been one of the driving forces of recent developments in network information theory. 
However, state-of-the-art coding schemes for TIM are usually handcrafted for specific families of network topologies, relying critically on experts’ domain knowledge and sophisticated treatments.
The lack of systematic and automatic generation of solutions inevitably restricts their potential wider applications to wireless communication systems, due to the limited generalizability of coding schemes to wider network configurations.
To address such an issue, this work makes the first attempt to advocate revisiting topological interference alignment (IA) from a novel learning-to-code perspective. Specifically, we recast the one-to-one and subspace IA conditions as vector assignment policies and propose a unifying learning-to-code on graphs (LCG) framework by leveraging graph neural networks (GNNs) for capturing topological structures and reinforcement learning (RL) for decision-making of IA beamforming vector assignment. Interestingly, the proposed LCG framework is capable of recovering known one-to-one scalar/vector IA solutions for a significantly wider range of network topologies, and more remarkably of discovering new subspace IA coding schemes for multiple-antenna cases that are challenging to be handcrafted. The extensive experiments demonstrate that the LCG framework is an effective way to automatically produce systematic coding solutions to the TIM instances with arbitrary network topologies, and at the same time, the underlying learning algorithm is efficient with respect to online inference time and possesses excellent generalizability and transferability for practical deployment.
\end{abstract}

\section{Introduction}
\label{sec:Introduction}

Topological interference management (TIM) is one of the most promising techniques for wireless networks with a much relaxed requirement of channel state information at the transmitters (CSIT). As introduced in \cite{jafar2013topological}, TIM examines the degrees of freedom (DoF) of partially connected one-hop wireless networks, with the only available CSIT being the network topology, i.e., a bipartite graph representing the connectivity between transmitters and receivers. Over the past few years, TIM has received extensive attention, resulting in a growing number of follow-up works, including TIM with alternating topology \cite{sun2013alternating,gherekhloo2013alternating2}, multi-level TIM \cite{geng2021multilevel}, multi-antenna TIM \cite{sun2014topologicalMIMO}, TIM with cooperation \cite{yi2015topological,yi2018topological}, dynamic TIM \cite{yi2019opportunistic,liang2022topological}, and many others (e.g., \cite{maleki2013optimality,yi2018tdma,doumiati2019framework,davoodi2018network,yang2017topological,aquilina2016degrees,shi2016low,gao2014topological,naderializadeh2014interference,mutangana2020topological}). 
Among various TIM coding schemes, interference alignment (IA) with topological CSIT, namely topological IA, 
has been proven to be information-theoretical optimal for a large family of network topologies \cite{jafar2013topological}.
It is worth noting that \textit{topological} IA is fundamentally different from the \textit{vanilla} IA \cite{8186780} that relies on global CSIT and extensive symbol extension, in such a way that topological IA goes far beyond the traditional designs of IA coding schemes, leveraging the only topological CSIT.

Albeit promising from a theoretical viewpoint, state-of-the-art topological IA coding schemes inspect each specific network topology individually and design transmit/receive precoding vectors in a handcrafted manner. This relies critically on experts' domain knowledge and delicate designs, which may be time-consuming, hard to generalize and scale to large-sized networks, restricting the potential applications to system designs. 
In particular, for the TIM setting with 6 transmitter-receiver pairs, there are 1.5 million instances or so with distinct topologies; the number of topologies increases exponentially as the number of pairs increases. It is extremely challenging, if not impossible, to handcraft every instance with an IA scheme, not to mention the much more complex multiple-antenna settings. It, therefore, calls for automatically generated systematic solutions to deal with the massive TIM instances and diverse multiple-antenna configurations.

As machine learning has been increasingly engaged in wireless communications, one may wonder if it can be leveraged to design and discover new coding techniques in an automatic and systematic way. Although some examples can be found in the coding literature (e.g., \cite{kim2020deepcode,chahine2021deepic,9517735} and many others), 
it is still an unexplored area as to how  
machine learning could be applied to coding on graphs for topological IA, where graph structures impose certain challenges.
In this paper, we make a first attempt to push forward this line of research, with the contributions summarized as follows:
\begin{itemize}
    \item 
Towards the TIM problem, we propose a novel intelligent coding framework for learning-to-code on graphs (LCG). The proposed LCG framework takes a special care of topological IA, 
translating beamformer design into dedicated vector generation followed by vector assignment according to IA conditions.
Specifically, LCG
leverages reinforcement learning (RL) for vector assignment, which employs graph neural networks (GNNs) as both the policy and value networks to capture topological structures.
As shown in Figure \ref{1st_framework}, for TIM instances, the directed message conflict graphs are first constructed from network topologies, followed by a learning-to-defer approach to assigning vectors to the conflict graph in an iterative manner with state transition until certain IA conditions are satisfied. Such a vector assignment strategy will be translated to topological IA for beamforming.

\item
By recasting topological IA to a vector assignment problem, we transform the one-to-one/subspace scalar/vector IA coding designs into vector assignment policies that can be learned with a sufficient number of unsolved TIM instances. In doing so, the proposed LCG framework is capable of recovering the majority of the one-to-one scalar/vector IA solutions known in the literature in a systematic way.
More remarkably, the proposed LCG framework is able to discover new subspace IA solutions for larger-sized networks and for the more challenging multiple-antenna configurations, which are difficult to design, or even imagine, in a handcrafted manner.
\begin{figure}[t]
\begin{center}
   {%
     \includegraphics[width=0.4\textwidth]{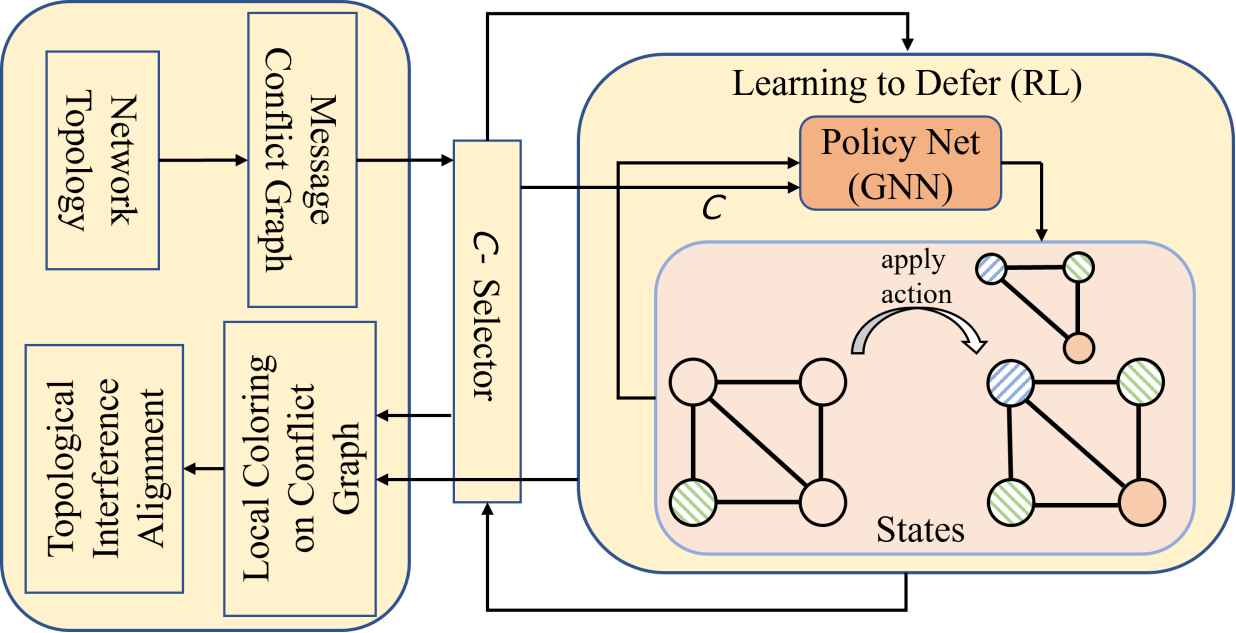}}
\end{center}
   \caption{The framework of the proposed learning-to-code on graphs (LCG), employing RL for learning to defer and GNNs for policy networks.
   }
   \label{1st_framework}
\end{figure}

\item
To evaluate the effectiveness of the proposed LCG framework, we conduct extensive experiments on various TIM instances with random network topologies.
For both synthetic and practical datasets (i.e., Erdős-Rényi random graphs and device-to-device networks), we found that the proposed LCG framework recovers the optimal coding solutions for the majority ($\ge 95\%$) of network topologies, and discovers subspace IA solutions for single-input multiple-output (SIMO) cases that outperform the single-input single-output (SISO) counterparts.
\end{itemize}

The rest of this paper is organized as follows. Section II provides the system model for the TIM problem and describes IA coding schemes. In Section III, we revisit topological IA from the vector assignment perspective, casting different one-to-one/subspace scalar/vector IA solutions into vector assignment conditions in the single-antenna and the multiple-antenna settings, and relate it to graph coloring. 
Section IV is dedicated to our proposed LCG framework, followed by some discussions in Section V on new structures discovered by LCG.
Finally, we detail experimental setups and evaluation results in Section VI, and conclude the paper in Section VII.

\section{System Model}
We consider the $K$-user interference channel model that has $K$ sources/transmitters, labeled as $S_1,S_2,\dots,S_K$, and $K$ destinations/receivers, labeled as $D_1,D_2,\dots,D_K$. We assume that each source is equipped $M$ antennas, and each destination is equipped $N$ antennas. At time instant $t$,  
Destination $D_j$ receives signal:
\begin{align}
Y_j(t) = \sum_{i=1}^K H_{ji}X_i(t)+Z_j(t),
\end{align}
where $X_i(t)\in \C^{M \times 1}$ is the symbol vector transmitted by Source $S_i$. $H_{ji}\in \C^{N \times M}$ is the channel coefficient matrix between Source $S_i$ and Destination $D_j$. $Z_j(t)\in \C^{N \times 1}$ is the additive white Gaussian noise (AWGN) at Destination $D_j$, and $Y_j(t)\in \C^{N\times 1}$ is the symbol received by Destination $D_j$. All transmitted signals are subject to a power constraint $P$, such that $\mathbb{E}(||X_i(t)||^2)\leq P$. Throughout this paper, we focus on the multiple unicast setting, where each source is paired with a unique destination with one desired message delivered. Without otherwise specified, the single-antenna case (i.e., $M=N=1$) is considered.

The TIM problem \cite{jafar2013topological} considers a partially connected interference network with the topology matrix $T = [t_{ji}]_{K\times K}$ known among all sources and destinations. Each element $t_{ji}$ is assigned a value of $0$ if a trivial channel exists from Source $S_i$ to Destination $D_j$, i.e., with weak capacity. In such case, $H_{ji}$ is set to $0$. Otherwise, $t_{ji} = 1$ if the channel is non-trivial.
\begin{figure}[htbp]
   \centering
   \vspace{-20pt}
   \subfloat[Network topology graph.]{%
     \label{5nodes_sample_topo}
     \includegraphics[width=3.1cm]{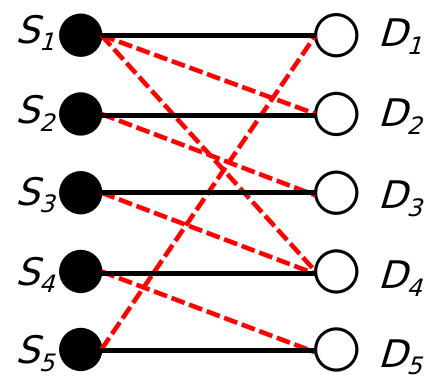}}
   \hfil
   \subfloat[Message conflict graph.]{%
     \label{5nodes_sample_conf}
     \includegraphics[width=3.4cm]{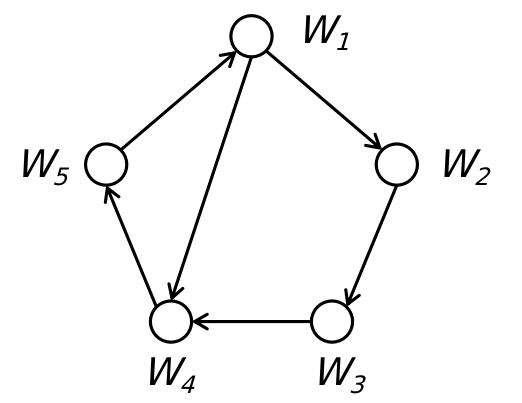}}  
   \caption{(a) The network topology graph of a 5-node TIM instance, where the black solid edges indicate paired sources and destinations with desired messages and the red dotted edges are interfering signals, and (b) the corresponding message conflict graph with desired messages (i.e., source-destination pairs) being nodes and the directed edges indicate interference from sources to destinations.}
   \label{5_nodes_sample}
\end{figure}
\subsection{TIM Representations}\label{TIM Representations}
Inspired by \cite{yi2018tdma}, we formally introduce two graph definitions for TIM representation.
\begin{defn}
Given the TIM problem with $K$-users, topology matrix $T$, and message set $\gM$, we define the following two graphs:
	\begin{itemize}
		\item [1)] 
		\textbf{Network Topology Graph}: An undirected bipartite graph
		with sources on one side, destinations on the other, and
		an edge between $S_i$ and $D_j$ whenever $t_{ji} = 1$.     
		\item [2)]
		\textbf{Message Conflict Graph}: A directed graph where each message $W_{i} \in \gM$ is a node, and a directed edge $(W_{i}, W_{j})$ exists if and only if $D_{j}$ is interfered by $S_i$, i.e., $t_{ji} = 1$, where $\gM$ is the set of desired messages.
	\end{itemize}
\end{defn}
One of the major objectives of the TIM problem is to maximize the symmetric DoF 
$$d_{\mathrm{sym}} = \min_{W_{i} \in \gM} \text{DoF}(W_i).$$
\begin{exe}
An example is shown in Figure \ref{5_nodes_sample}. Note that a connection between a source $S$ and a destination $D$ could be one of two cases: demanded link and interfering link. The link is demanded if there exists a desired message from $S$ to $D$ (black solid lines in Figure \ref{5nodes_sample_topo}) and interfering otherwise (red dotted lines in Figure \ref{5nodes_sample_topo}). A similar undirected version of message conflict graph was defined in \cite{yi2018tdma}, which ignores the direction of interference. \hfill $\square$
\end{exe}

\subsection{Linear Coding Schemes and Outer Bounds}
The majority of TIM solutions are linear coding schemes \cite{jafar2013topological}. A linear coding scheme over $C$ channel uses achieving the DoF
$$\text{DoF}(W) = \dfrac{b}{C}, \quad \forall W \in \gM$$
where $b$ is a non-negative integer, consists of
\begin{itemize}
    \item[1)] precoding matrices of Source $S_i$, $\mathbf{V}_i \in \C^{C\times b}, i\in \{1,2,\dots, K\}$, and
    \item[2)] combining matrices of Destination $D_i$, $\textbf{U}_i\in \C^{b\times C}, i\in \{1,2,\dots , K\},$
\end{itemize}
such that the following properties are satisfied: 
\begin{IEEEeqnarray}{rll}
&\mathbf{U}_j \mathbf{V}_i=0, \forall i,j\in \{1,2,\dots , K\}, i \neq j \text{ and }t_{ji} = 1,\\
&\det(\mathbf{U}_j\mathbf{V}_j)\neq 0,\forall j \in \{1,2,\dots , K\}.
\end{IEEEeqnarray}

Each message is split into $b$ independent scalar streams, each of which carries one symbol from $\C$, and is transmitted
along the corresponding column vectors (the ``beamforming'' vectors) of the precoding matrix $\mathbf{V}$. Once the precoding matrix $\mathbf{V}$ is fixed, the combining matrix can be obtained through the zero-forcing method. Therefore, in what follows, we concentrate on the design of $\mathbf{V}$, leaving zero-forcing matrices $\mathbf{U}$ aside. According to the value of $b=1$ or $b>1$, we divide linear coding into scalar and vector schemes, respectively. Note that ``scalar'' refers to each message being associated with a single symbol, which is still transmitted along vector.

According to \cite{jafar2013topological}, any outer bound of an index coding problem with the side information graph is also an outer bound of the corresponding single-antenna TIM instance with message conflict graph being the complement of the side information graph.
Therefore, we can employ well-established outer bounds for index coding, such as the maximal acyclic induced subgraph (MAIS) bound \cite{bar2011index} and the minimum internal conflict distance (MICD) bound \cite{jafar2013topological}, to validate the optimality of the linear coding schemes. In the later experiments, we use the MAIS bound, whose definition is as follows, to assess the optimality of the proposed model.
{
\begin{lem}[MAIS bound]
The symmetric DoF of any TIM with message conflict graph $\gG$ is at most $\frac{1}{\textsc{MAIS}(\gG_c)}$, where \textsc{MAIS}($\gG_c$) is the size of the maximum acyclic induced subgraph of $\gG_c$, and $\gG_c$ is the complement of $\gG$.
\end{lem}
}

\section{Revisiting Interference Alignment}\label{Interference Alignment Perspective}

Among many linear coding schemes for TIM, IA is one of the most promising approaches. IA aims to align signals from interfering transmitters as much as possible, while maintaining the signal from the desired transmitter separated. 
Based on how interfering messages are aligned, IA can be categorized into two classes: one-to-one IA and subspace IA. 

One-to-one IA involves the perfect alignment of interferences in a one-to-one manner, where the beamforming vectors of aligned interfering symbols are identical, in such a way that the interfering signals are linearly dependent with the desired signals in the destination's viewpoint.
Subspace IA, on the other hand, is to align interferences in a subspace with reduced dimensions.
In this case, the beamforming vectors carrying interfering symbols are not required to perfectly align one another on a one-to-one basis, but one interfering symbol can align itself in the subspace spanned jointly by the beamforming vectors of other interfering symbols.

By combining scalar/vector linear coding and one-to-one/subspace IA, we obtain four types of IA coding schemes: one-to-one scalar IA (OSIA), one-to-one vector IA (OVIA), subspace scalar IA (SSIA), and subspace vector IA (SVIA). It is noteworthy that there exist inclusion relationships among their solution spaces. Specifically, 
\begin{IEEEeqnarray}{rll} \label{eq:IA-relation1}
\text{OSIA}&\subseteq \text{OVIA} \subseteq \text{SVIA} , \text{and}\\ \label{eq:IA-relation2}
 \text{OSIA}&\subseteq \text{SSIA} \subseteq \text{SVIA}.
\end{IEEEeqnarray}

Interestingly, IA can be captured by the directed message conflict graph. We define the in-neighborhood of node $j$, denoted as $\mathcal{N}^+(j)$, to be the set of all nodes pointing to $j$. The closed in-neighborhood $\mathcal{N}_c^{+}(j)$ of node $j$ is the union of $j$ and $j$'s in-neighborhood, i.e., $\mathcal{N}_c^{+}(j) = \{j\} \cup \mathcal{N}^+(j)$.

Let us suppose that the beamforming matrix $\mathbf{V}_i$ is assigned to the Source $S_i$, or equivalently, to node $i$. The interference space of Destination $D_j$, or equivalently, of node $j$, is denoted by $\I_j$ and is defined as follows:
\begin{IEEEeqnarray}{rCl}
\I_j &= \operatorname{span} (\oplus_{i \in \mathcal{N}^+(j)} (H_{ji}\otimes \mathbf{V}_i)),
\end{IEEEeqnarray}
where $\oplus$ represents the matrix concatenation along the first dimension, and $\otimes$ denotes the Kronecker product operation.

Similarly, the overall signal space, i.e., the union of desired signal and interference space, of node $j$ is denoted by $\S_j$ and is defined as:
\begin{IEEEeqnarray}{rCl}
\S_j &= \operatorname{span} (\oplus_{i \in \mathcal{N}_c^+(j)} (H_{ji}\otimes \mathbf{V}_i)).
\end{IEEEeqnarray}

We then formally define the linear coding space and the IA task.
\begin{defn}
The linear coding space, denoted as $\mathcal{B}(C)$, is defined as the space spanned by $C$ arbitrary linearly independent basis vectors of dimension $C \times 1$.
\end{defn}

{
An IA coding scheme on a TIM message conflict graph $\gG$
can be implemented by assigning appropriate beamforming matrices $\mathbf{V}_i$ to Sources $S_i$. 
To achieve a symmetric DoF value of $d_{\mathrm{sym}} = \frac{b}{C}$, the IA design can be formulated explicitly as the following optimization problem:
\begin{IEEEeqnarray}{lll}
\underset{\mathbf{V}}{\text{maximize}} \quad & d_{\mathrm{sym}} = \frac{b}{C} \\\label{IAcondition1}
\text{subject to} \quad & \mathrm{rank}(\S_j) - \mathrm{rank}(\I_j) = b, \ \forall j \in \gV, \\\label{IAcondition2}
& \mathbf{V}_i \in \mathbb{C}^{C \times b} \cap \mathcal{B}(C), \  \forall i \in \{1,\dots,K\},\\\label{IAcondition3}
&b \leq C. 
\end{IEEEeqnarray}
Here, the rank of a space refers to the maximum number of linearly independent vectors within that space. 
}
\begin{lem}
An IA coding scheme satisfying Eq-(\ref{IAcondition1})-(\ref{IAcondition3}) achieves a symmetric DoF value given by
\begin{equation}
    d_{\mathrm{sym}} = \frac{b}{C}.
\end{equation}
\end{lem}

\begin{IEEEproof}
We prove that such an IA coding scheme is decodable on each destination. 
Let's denote the desired signal space of Destination $D_j$ to be $\D_j = \operatorname{span}(H_{jj}\otimes \mathbf{V}_j).$ 
By the definition, we have $\S_j = \D_j \cup \I_j.$
Therefore,
\begin{IEEEeqnarray*}{lll}
    &&\ \mathrm{rank}(\S_j)-\mathrm{rank}(\I_j)\\
    &=&\ \mathrm{rank}(\D_j)+\mathrm{rank}(\I_j)-\mathrm{rank}(\D_j \cap \I_j) - \mathrm{rank}(\I_j)\\
    &=&\ \mathrm{rank}(\D_j)-\mathrm{rank}(\D_j \cap \I_j)\\ 
    &=&\ b. 
\end{IEEEeqnarray*}
Since $\mathrm{rank}(\D_j) \leq b$ naturally, we have
$$\mathrm{rank}(\D_j) = b, \text{and}\ \mathrm{rank}(\D_j \cap \I_j)=0,\forall j \in \gV,$$
i.e., $\D_j \cap \I_j = \emptyset.$

That is to say, the desired signal space of each destination has full-rank, and the desired signal space is not interfered by the interference space. Hence, each destination would be able to decode the desired message. 
\end{IEEEproof}

Suppose $b$ is fixed and $C$ is variable. We now claim that the key to increasing the achieved DoF value is to reduce the maximum rank of $\I_j$ over all $j \in \gV$.
Since the achieved DoF value is $\frac{b}{C}$, we aim to reduce $C$. However, it is essential that $C$ satisfies $C \geq \max_{j \in \gV} \mathrm{rank}(\S_j)$ because $\S_j \subseteq \mathcal{B}(C)$. Therefore, it is suitable to set $$C = \max_{j \in \gV} \mathrm{rank}(\S_j) = b+\max_{j \in \gV} \mathrm{rank}(\I_j).$$

Thus, reducing the maximum rank of $\I_j$ is essential to increase the achievable DoF value. Following this, we will proceed to introduce IA within the contexts of single-input single-output (SISO), single-input multiple-output (SIMO), and  multiple-input single-output (MISO) configurations, respectively.

\subsection{SISO TIM IA}
Consider SISO networks where each source/destination is equipped with only one antenna, i.e. $M=N=1$. This implies that for Destination $D_j$, $H_{ji} = h_{ji} \in \C$ and $H_{ji} \otimes \mathbf{V}_i$ degenerates to $h_{ji}\mathbf{V}_i$. The dimensions of the interference space $\dim(\mathcal{I}_j)$ and that of signal space $\dim(\mathcal{S}_j)$ degenerate to $\dim(\operatorname{span} (\oplus_{i \in \mathcal{N}^+(j)} \mathbf{V}_j))$ and $\dim(\operatorname{span} (\oplus_{i \in \mathcal{N}_c^+(j)} \mathbf{V}_i))$ as well. This implies that symbols transmitted through linearly dependent beamforming vectors remain dependent within the signal space, and vice versa.
We then discuss the four types of IA coding schemes with examples.
\subsubsection{One-to-One Scalar Interference Alignment (OSIA)}
Each source transmits a single symbol\footnote{The terminologies of symbol and message are used interchangeably for ease of presentation when referring to the scalar case.} ($b=1$), with a corresponding beamforming vector in such a way that each interfering signal is perfectly aligned with another one in the receiver's signal space, leaving one interference-free dimension for each desired signal. By aligning interfering signals within a reduced subspace of the receiver's signal space, OSIA can improve the achievable DoF over orthogonal access.

Therefore, the coding scheme design for scalar IA can be conducted by first generating a linear coding space $\mathcal{B}(C)$, and then assigning a beamforming vector $\mathbf{V}_{C\times 1} \in \mathcal{B}(C)$ to each node (i.e., message) in the message conflict graph.
To achieve SISO OSIA, the vector assignment should meet the following two conditions:
\begin{itemize}
  \item [\textbf{C1})] Connected nodes should be assigned linearly independent vectors, regardless of the direction of edges.
  \item [\textbf{C2})] Nodes pointing to the same node should be assigned as few different vectors as possible.
\end{itemize}

\begin{figure}
   \centering
   \subfloat[Ex.~2]{%
     \label{Ex. 2: OSIA}
     \includegraphics[width=3.3cm]{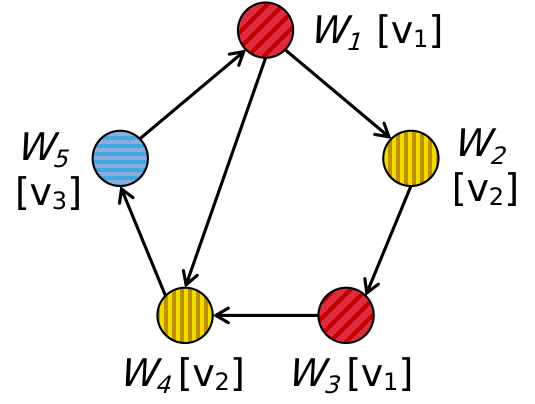}}
   \hfil
   \subfloat[Ex.~4]{%
     \label{Ex. 4: OVIA}
     \includegraphics[width=3.8cm]{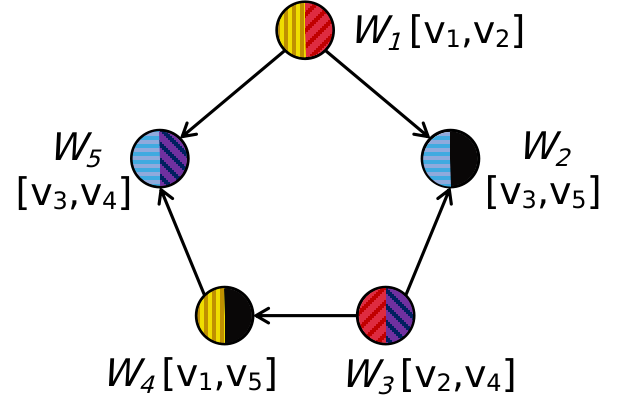}}  
   \caption{(a) Conflict graph of Ex.~2 and a OSIA solution as local coloring achieving optimal DoF 1/2, and (b) conflict graph of Ex.~4 and a OVIA solution as fractional (local) coloring achieving optimal DoF 2/5.}
   \label{5_nodes_sample_vectorIA}
\end{figure}

\begin{exe}
As shown in Figure \ref{Ex. 2: OSIA}, OSIA can be interpreted as follows. The message $W_{4}$ sees two incoming edges from $W_{1}$ and $W_{3}$. If both messages $W_{1}$ and $W_{3}$ live in a subspace that does not contain $W_{4}$, then the desired message is separable from the interfering ones. Thus, we can generate a 2-dim linear coding space $\mathcal{B}(2)$ spanned by $\v_1=[1 \; 0]^T$ and $\v_2=[0 \; 1]^T$. Then, the vector assigned to the messages $W_{1}$, $W_{2}$, $W_{3}$, $W_{4}$, and $W_{5}$ will be $\v_1$, $\v_2$, $\v_1$, $\v_3=\v_1+\v_2$, and $\v_2$, respectively. Note here that, any pair of connected nodes, e.g. $W_{4}$ and $W_{5}$, are assigned linearly independent vectors, and the messages (e.g., $W_{1}$ and $W_{3}$) pointing to the same message $W_{4}$ are assigned the same vector for alignment. This yields an achievable DoF value of $b/C=1/2$, which is also optimal, according to the MAIS outer bound. \hfill $\square$
\end{exe}

\textit{The relationship with graph local coloring}:
The relationship between the TIM problem and graph vertex coloring problem has been investigated through index coding by \cite{shanmugam2013local}. 
When two vectors are chosen from the same set of basis vectors, they are either identical (with inner product of 1) or orthogonal (with inner product of 0). It is analogous to vertex coloring on two endpoints of an edge - the identical color leads to a conflict (i.e., `1') and distinct colors yield an independent set (i.e., `0'). As we consider the DoF metric, the identical/orthogonal constraints can be relaxed to dependence/independence.
If we interpret linear independent vectors as different colors, then vector assignment can be alternatively done by color assignment. Vertex coloring aims to color the nodes of an undirected graph such that no two connected nodes are assigned the same color. This agrees with \textbf{C1} of IA. Local vertex coloring is a special variant of vertex coloring on directed graphs, which was first introduced in \cite{KORNER2005101}.
It aims to color a directed graph, while minimizing the maximum number of colors used in each local in-neighborhood. This agrees with \textbf{C2} of IA.

Given a TIM instance, the traditional handcrafted method \cite{jafar2013topological,sun2014topologicalMIMO, maleki2014index} typically begins by analyzing the theoretical outer bound of DoF on a case-by-case basis, without precise knowledge of the beamforming vectors. Then, a set of beamforming vectors is carefully designed to achieve this bound. In contrast, our proposed LCG approach takes the opposite direction. LCG first creates the linear coding space $\mathcal{B}(C)$, and generates a sufficient number of vectors in $\mathcal{B}(C)$, such that any $C$ of them are linearly independent. These vectors are then used as beamforming vectors to automatically solve the IA problem, or its equivalent graph coloring problem, on the message conflict graph. Suppose a local coloring scheme where the maximum number of colors in each closed in-neighborhood is $C$ is obtained. One can arbitrarily establish a mapping between vectors and colors, and assign vectors to nodes according to the local coloring scheme. It can be proved that the vector assigned to any node $j$ is \textit{distinct} and \textit{linearly independent} from the vectors assigned to $j$'s in-neighborhood. The distinction comes from the rule of vertex coloring, and the linear independence comes from the maximum number of colors in each closed in-neighborhood is $C$, and every $C$ vectors being linearly independent. Therefore, all destinations will be able to decode the demanded messages, as the interfering signals are perfectly aligned with a $(C-1)$-dim subspace, leaving a 1-dim space clean to recover the desired message. 
\begin{figure}[htbp]
	\begin{center}
		\includegraphics[width=0.4\textwidth]{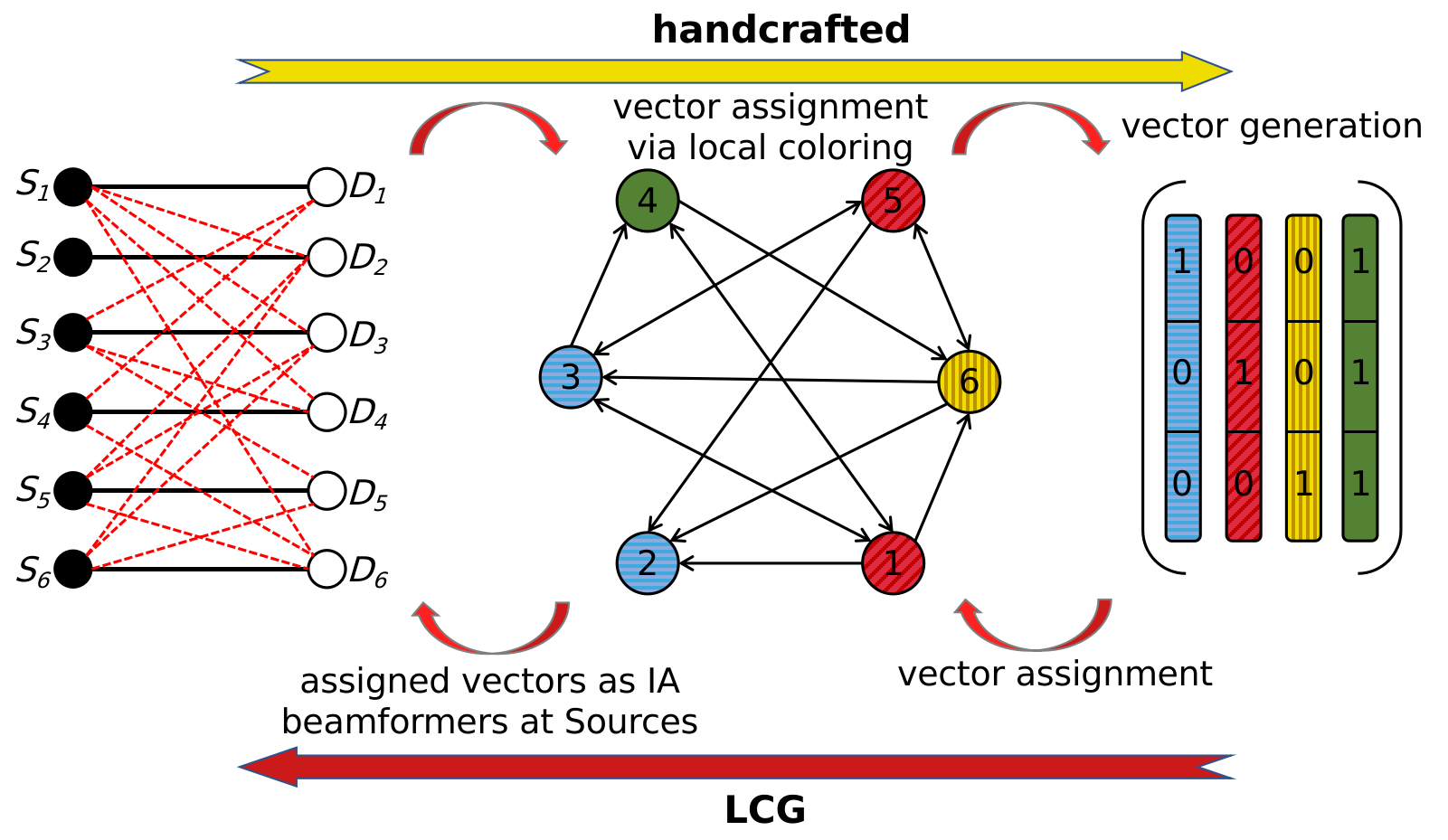}
	\end{center}
	\caption{Ex. 3: an example of solving TIM instance through local coloring. Comparison between the handcrafted method and LCG.}
	\label{MDS_exaple}
\end{figure}
\begin{exe}
An example of this process is shown in Figure \ref{MDS_exaple}. We first create the linear coding space $\mathcal{B}(C)$ with $C=3$ (the setting of parameter $C$ is explained in Section \ref{overview_LCG}), and generate $4$ beamforming vectors (or colors) as illustrated on the right-hand side of the diagram. Then LCG uses these vectors or colors to solve the IA problem or graph coloring problem. In this example, nodes $1$ and $5$ are assigned the red color and are aligned together in the signal space of nodes $2,3$ and $6$, resulting in the achievable DoF value of $1/3$, which is also optimal according to the MAIS outer bound. \hfill $\square$
\end{exe}

\subsubsection{One-to-One Vector Interference Alignment (OVIA)}
Each source is allowed to transmit multiple symbols ($b>1$) via several beamforming vectors over time, whereas the interfering signals are still aligned in a one-to-one manner. 
Different from the scalar case where the signals from different transmitters are either perfectly aligned or linearly independent, the vector case allows for partial alignment of interfering signals from different transmitters.
This enhances the flexibility of the design of interference alignment, leading to increased symmetric DoF compared to OSIA in some scenarios \cite{roth2006introduction}. 

\begin{exe}\label{Ex_OVIA}
Figure \ref{Ex. 4: OVIA} depicts the message conflict graph of a TIM instance, where the use of OSIA results in a symmetric DoF of 1/3, which is equivalent to what can be achieved with orthogonal access such as time-division multiple access (TDMA). However, using the OVIA scheme, as demonstrated subsequently, we can increase the achievable DoF value to 2/5.
The message $W_{5}$ sees two incoming edges from $W_{1}$ and $W_{4}$. The message $W_{2}$ sees two incoming edges from $W_{1}$ and $W_{3}$. If the messages $W_{1}$, $W_{4}$ live in the subspace that does not contain $W_{5}$, and the messages $W_{1}$, $W_{3}$ live in the subspace that does not contain $W_{2}$ then the desired message $W_{5}$ and $W_{2}$ could be separable from the interfering ones.

Hence, the coding scheme design for vector IA involves generating a linear coding space $\mathcal{B}(C)$ and assigning a beamforming matrix $\mathbf{V}_{C\times b} \in \mathcal{B}(C)$ to each node in the message conflict graph. The vector assignment should follow the same conditions, \textbf{C1} and \textbf{C2}, if we consider $\mathbf{V}_{C\times b}$ as comprising $b$ vectors with dimension $C\times 1$.
Hence, the OVIA coding scheme can be obtained as follows.
By randomly generating 5 vectors $\{\v_1,\v_2,\dots,\v_5\}$ from $\mathcal{B}(5)$, that are in general linearly independent position, over a sufficiently large field. We assign beamforming matrix as shown in Figure \ref{Ex. 4: OVIA}.
Each message is split into two symbols, each of which is transmitted by some beamforming vector $\v$.
As a result, at the receiver side, the interfering signals occupy at most 3-dim subspace, e.g., 
$\dim([\v_1,\v_2],[\v_1,\v_5])=\dim([\v_1,\v_2],[\v_2,\v_4])=3$,
effectively aligning interference
and leaving at least a 2-dim interference-free subspace for the desired signal. This yields $d_{\mathrm{sym}}=2/5$ achievable, which is also optimal according to the MICD bound.    \hfill $\square$
\end{exe}

\textit{The relationship with graph fractional local coloring:}
In fractional coloring problem  (or $b$-fold coloring) \cite{scheinerman2011fractional}, each node is assigned $b$ colors rather than just one color, such that no two connected nodes share any common colors. Fractional local coloring is a fractional version of local coloring, aiming to minimize the maximum number of colors used in each in-neighborhood for any valid fractional coloring. If we regard the $b$ colors assigned to a node as the $b$ linear independent beamforming vectors, then the requirements of fractional local coloring perfectly match those of OVIA. Ensuring that any two connected nodes share no common colors guarantees condition \textbf{C1}, while minimizing the maximum number of colors in each in-neighborhood agrees with condition \textbf{C2}.

With such a fractional local color assignment, we can proceed to generate beamforming vectors as previously described. Subsequently, we allocate $b$ vectors to each node of the message conflict graph, thereby achieving a valid OVIA coding scheme.

\begin{figure}
   \centering
   \subfloat[Ex.~5: Network topology.]{%
     \label{4nodes_sample_topo_subspacescalar}
     \includegraphics[width=3.1cm]{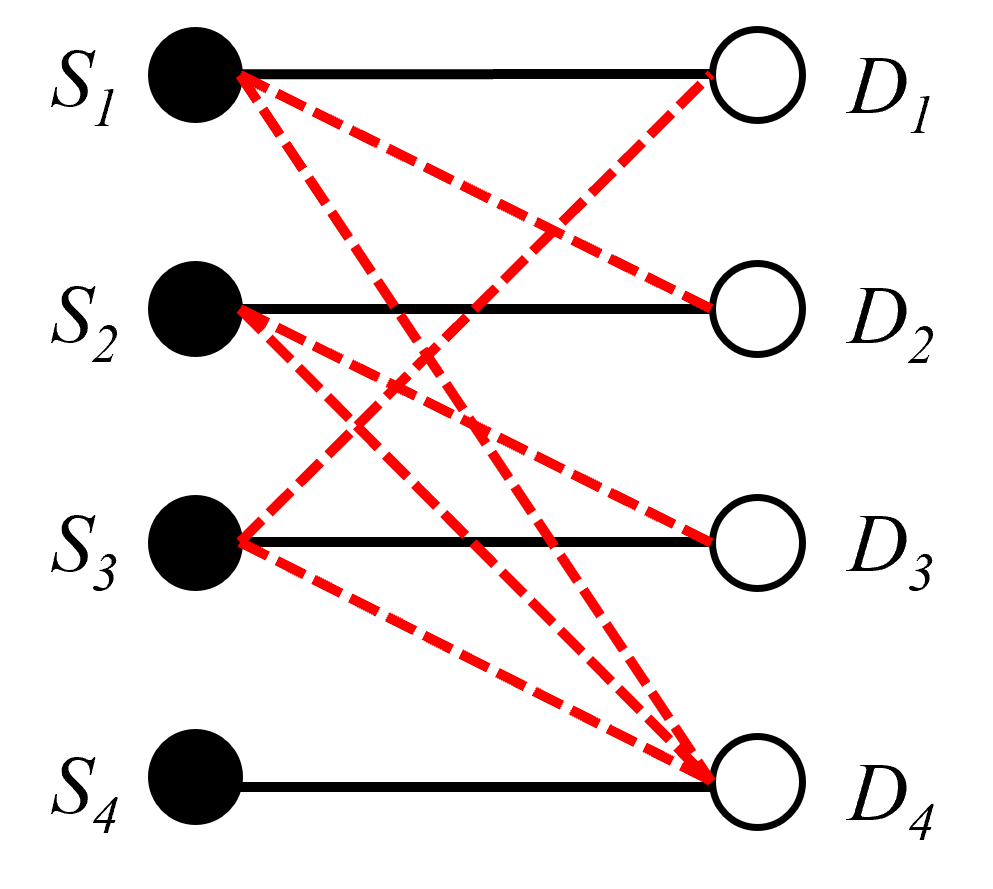}} 
   \hfil
   \subfloat[Ex.~5: Conflict graph.]{%
     \label{Ex. 5 SSIA}
     \includegraphics[width=3.9cm]{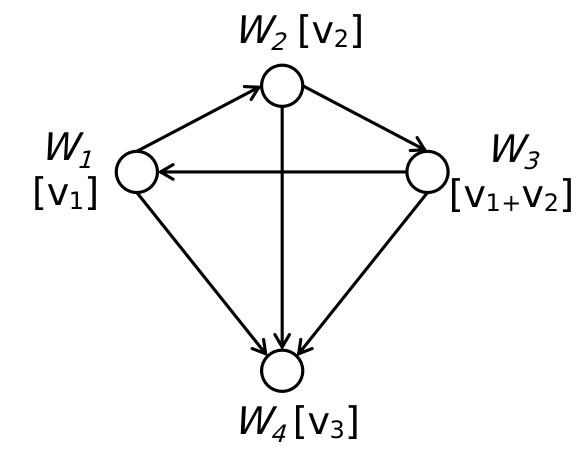}}  
   \caption{(a) The TIM instance topology graph of Ex.~5, and (b) a subspace scalar IA solution of Ex.~5 achieving optimal DoF 1/3.}
   \label{4_nodes_sample_subspace_scalar}
\end{figure}

\subsubsection{Subspace Scalar Interference Alignment (SSIA)}
Going beyond one-to-one IA, SSIA aims to align interference from multiple transmitters into a subspace of the receiver's signal space while maintaining a separate interference-free subspace for the desired signal. In SSIA, interfering signals do not need to align perfectly with each other but are confined to a lower-dimensional subspace formed by other interfering signals. This enables the receiver to isolate and remove interfering signals within that subspace, while the desired transmitter operates independently in a separate interference-free subspace

In certain scenarios, SSIA can increase the symmetric DoF of TIM over OSIA. 
By aligning the interfering signals to a subspace of the receiver's signal space, SSIA can effectively suppress interference from multiple sources, allowing for the simultaneous transmission of more independent data streams. The vector assignment should follow the following two conditions to meet SISO subspace IA:
\begin{itemize}
    \item [\textbf{C3})] Vectors assigned to each node can not belong to the subspace spanned by the vectors assigned to its in-neighborhood.
    \item [\textbf{C4})] Vectors assigned to the in-neighborhood of each node should occupy as small dimensional subspace as possible. 
\end{itemize}
\begin{exe} \label{Ex_SVIA}
As presented in Figure \ref{4_nodes_sample_subspace_scalar}, serves as an illustrative example, where both OSIA and OVIA fail to achieve the DoF value beyond $1/4$, while SSIA achieves the DoF value of $1/3$. Specifically, the messages $W_{1}$, $W_{2}$, and $W_{3}$ all point to (interfere with) $W_{4}$, therefore they need to be aligned as much as possible. At the same time, these three messages form a directed cycle, indicating that they also need to be distinguished from each other pairwise. In this case, one of the messages, for example $W_{3}$, can be aligned with the subspace spanned by $W_{1}$ and $W_{2}$, while ensuring that they are linearly independent from each other. Meanwhile, $W_{4}$ must not fall within the subspace formed by $W_{1}$, $W_{2}$, and $W_{3}$, allowing for the separation of desired message from the interfering ones.

For instance, we can generate a 3-dim linear coding space $\mathcal{B}(3)$ spanned by three linearly independent vectors $\v_1=[1 \; 0\; 0]^T$, $\v_2=[0 \; 1\; 0]^T$ and $\v_3=[0 \; 0 \; 1]^T$. Then, we assign beamforming vectors $\v_1$, $\v_2$, $\v_1+\v_2$, and $\v_3$ to the messages $W_{1}$, $W_{2}$, $W_{3}$, and $W_{4}$, respectively, as shown in Figure \ref{Ex. 5 SSIA}. Note here that, the dimension of the interference subspace spanned by the vectors assigned to the in-neighborhood of $W_{4}$ is 2, i.e., $\dim(\operatorname{span}(\v_1, \v_2, \v_1+\v_2) = 2$, leaving 1-dim interference-free subspace to $W_{4}$. Meanwhile, $W_{1}$, $W_{2}$, $W_{3}$ are pairwise separable, as any two of their beamforming vectors $\v_1$, $\v_2$, and $\v_1+\v_2$ are linearly independent. This yields an achievable symmetric DoF of $1/3$, which is also optimal according to the MAIS bound.\footnote{Since the graph coloring is not adequate to specify the overlap of different subspaces (cf. partial subspace alignment), the mapping between one-to-one IA and graph coloring is not applicable in subspace IA. }   \hfill $\square$ 
\end{exe}

\subsubsection{Subspace Vector Interference Alignment (SVIA)}
When multiple symbols are transmitted from each source with a precoding matrix, SVIA 
seeks to align the interference from multiple interfering transmitters to a specific subspace within the receiver's signal space, while ensuring that a subspace orthogonal or independent of the interference subspace is preserved for the desired signals.

Specifically, let $b$ symbols be sent from each source with a $C \times b$ beamforming matrix. Similar to SSIA, SVIA does not require the $b$ column vectors of the beamforming matrix to be perfectly aligned with those from other sources in a one-to-one manner, but rather the subspace spanned by the column vectors is considered such that the interfering signals are forced to live in a subspace with reduced dimensionality.
As the only difference between SSIA and SVIA is the number of symbols each user could send, SVIA can be simply implemented by repeating SSIA multiple times with carefully designed beamforming vectors. An example is given in Section V-a.

\subsection{SIMO TIM IA}
The symmetric MIMO TIM, where all nodes have the same number of antennas, i.e., $M=N>1$, is examined in \cite{jafar2013topological}. It still essentially represents the index coding problem, and the achievable DoF region through linear schemes includes the scaled version of the SISO setting, satisfying the spatial scale invariance property. Moreover, symmetric MISO TIM, where all sources are equipped with multiple antennas and all destinations are equipped with single antennas, i.e., $M>1,N=1$, is studied in \cite{sun2014topologicalMIMO}. It has been verified that solely increasing the number of transmit antennas does not enhance the DoF. However, the symmetric SIMO TIM, where all sources are equipped with single antenna and all destinations are equipped with multiple antennas, i.e., $M=1,N>1$, remains open and poses significant challenges. Therefore, in this section, we delve into exploring the symmetric SIMO TIM problem.

We now explain the changes brought about by SIMO compared to SISO and its significant contribution to improving DoF. The core change brought by the SIMO network is the per-antenna separability constraint, i.e., the aligned interference should be separable from the desired signal over \textit{each} receive antenna \cite{sun2014topologicalMIMO}.
Specifically, each destination is capable of tolerating $N-1$ interferences across each dimension of the linear coding space. This is due to the fact that $H_{ji} \otimes [\v_1, \v_2, \dots, \v_{N}]$ has full rank almost surely, where $H_{ji}\in \C^{N\times 1}$ and $\v_1, \v_2, \dots, \v_{N}$ represent $N$ randomly generated vectors with dimension $C\times 1$ such that $C \geq N$, even if $\v$s are pairwise linearly dependent.  In contrast, in a SISO network, the desired transmitter must transmit a signal using a beamforming vector that is independent of the beamforming vectors of all interfering signals. Otherwise, the desired signal will become entangled with the interference, rendering it undecodable. 

Another change is that one-to-one IA is usually infeasible in the SIMO setting, 
because even two identical beamforming vectors, say $\v_1=\v_2$, will be unnecessarily aligned from the viewpoint of the destination, i.e., $\operatorname{span}(H_{ji}\otimes \v_1, H_{ji'}\otimes \v_2)$ usually
spans a subspace almost surely. Therefore, we suppose the alignment of interferences in a subspace manner.

The vector assignment should follow the following two conditions to meet SIMO IA:
\begin{adjustwidth}{1em}{}
\begin{itemize}
  \item [\textbf{C5})] For \textit{each} vector assigned to a message (node), there should be at most $N-1$ other nodes assigned with the same vector that point to such node.
  \item [\textbf{C6})] The dimension of the interference subspace spanned by the assigned vectors within the in-neighborhood of any node should be as small as possible.
\end{itemize}
\end{adjustwidth}
We denote the SIMO network with $N$ antennas at each destination as SIMO-$(1,N)$. While certain understanding is available in the literature (e.g., \cite{sun2014topologicalMIMO}), the SIMO TIM problem is still largely unexplored. 
\begin{figure}[t]
   \centering
   \subfloat[Ex.~6: SISO.]{%
     \label{MIMO_example1_OSIA}
     \includegraphics[width=3.5cm]{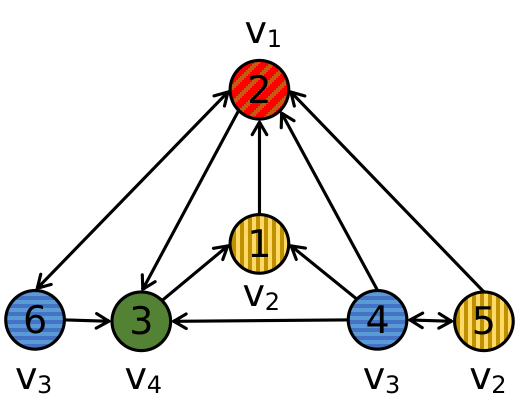}}  
   \hfil
   \subfloat[Ex.~6: SIMO-$(1,2)$.]{%
     \label{MIMO_example1_MIMO}
     \includegraphics[width=3.5cm]{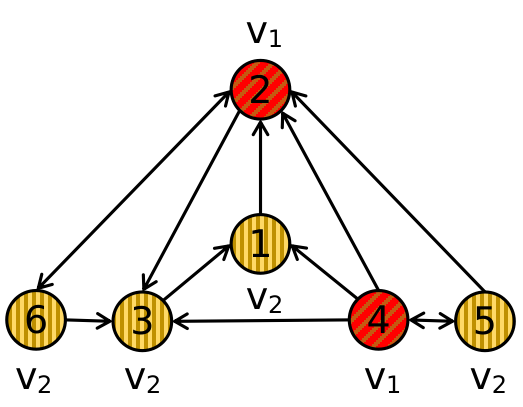}}  
   \caption{(a) A SISO TIM instance and a one-to-one scalar IA solution achieving optimal DoF 1/3, and (b) a SIMO-(1,2) TIM instance and a scalar IA solution achieving DoF 1/2.}
   \label{MIMO_example1}
\end{figure}

\begin{figure*}[t]
   \centering
   \subfloat[Ex.~\ref{large_graph_exe}: SISO.]{%
     \label{large_graph_anttena1}
     \includegraphics[width=4.5cm]{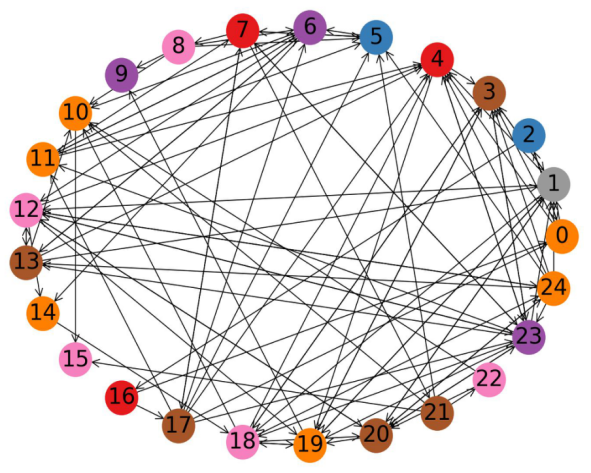}}  
   \hfil
      \subfloat[Ex.~\ref{large_graph_exe}: SIMO-$(1,2)$.]{%
     \label{large_graph_anttena2}
     \includegraphics[width=4.5cm]{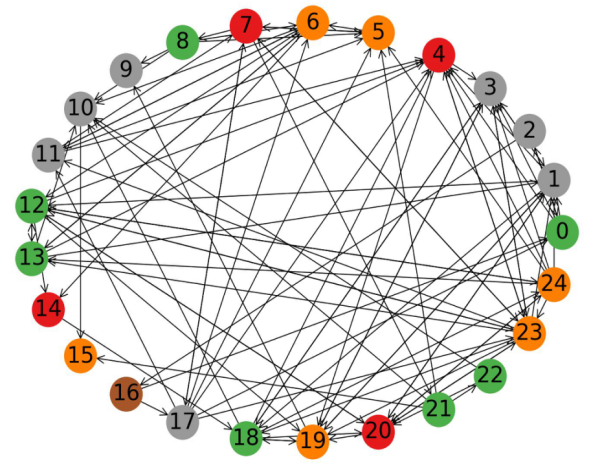}}  
   \hfil
   \subfloat[Ex.~\ref{large_graph_exe}: SIMO-$(1,3)$.]{%
     \label{large_graph_anttena3}
     \includegraphics[width=4.5cm]{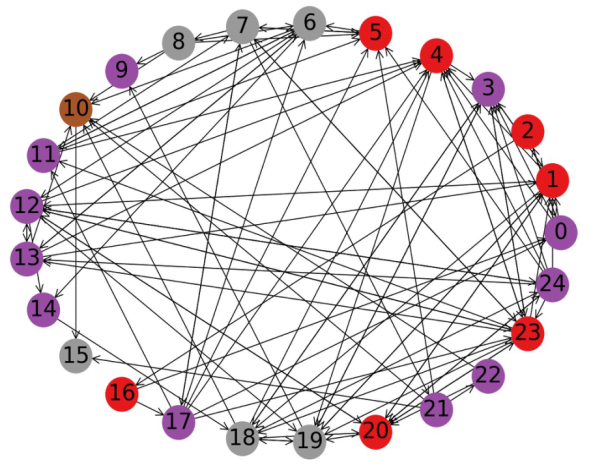}}  
   \caption{(a) A SISO OSIA IA solution achieving DoF 1/6, (b) a SIMO-$(1,2)$ scalar IA solution achieving DoF 1/4 and, (c) a SIMO-$(1,3)$ scalar IA solution achieving DoF 1/3.}
   \label{large_graph_figure}
\end{figure*}
\begin{exe}
Figure \ref{MIMO_example1} shows an example that achieves better DoF value under the SIMO setting. First, when the SISO case is considered, it can be observed that orthogonal methods (e.g., TDMA) achieve the DoF value of $1/4$ since there exists a fully connected subgraph with $4$ nodes (also called $4$-clique). 

Figure \ref{MIMO_example1_OSIA} shows a OSIA scheme that achieves the DoF value of $1/3$, which is optimal according to the MAIS bound. We can choose the vectors as follows 
$\v_1=[1 \; 0\; 0]^T, \v_2=[0 \; 1\; 0]^T, \v_3=[0 \; 0 \; 1]^T \text{ and } \v_4=[1 \; 1\; 1]^T$, 
and assign them to nodes as illustrated in Figure \ref{MIMO_example1_OSIA}.

However, when it comes to a SIMO-$(1,2)$ network, a scalar IA scheme achieves the improved DoF value of $1/2$, as shown in Figure \ref{MIMO_example1_MIMO}. One can validate the achievability by checking that: 1) each node is pointed to by at most $2-1=1$ node with the same color as itself, so that the carrying messages can be decoded with two receive antennas using e.g., zero-forcing combining; and 2) there are at most $2$ colors within each closed in-neighborhood, so that the interferences are perfectly aligned. Specifically, we could choose $\v_1=[1 \; 0]^T$ and $\v_2=[0 \; 1]^T$, and assign them as in Figure \ref{MIMO_example1_MIMO} to obtain DoF of 1/2. \hfill $\square$
\end{exe}

\begin{exe}\label{large_graph_exe}
The instance in Figure \ref{large_graph_figure} has 25 nodes, making it difficult to manually design even the simplest OSIA schemes. We consider the symmetric SIMO settings with three different antenna configurations, where all destinations have 1, 2, and 3 antennas, i.e., $N=1, 2, 3$, respectively. By using LCG to produce the linear coding schemes as shown in Figure \ref{large_graph_anttena1}, \ref{large_graph_anttena2}, and \ref{large_graph_anttena3}, where distinct colors indicate different beamforming vectors, we achieve the corresponding DoF values $1/6$, $1/4$, and $1/3$, respectively. An important note is that it only takes the LCG framework approximately 0.2 seconds to produce each coding scheme with an RTX 4060 GPU. This is in sharp contrast to the handcrafted approach. \hfill $\square$
\end{exe}

\section{Learning to Code on Graphs}
From the aforementioned IA coding schemes, it appears that various IA coding can be implemented by properly generating beamforming vectors and assigning them to different nodes with certain conditions \textbf{C1}-\textbf{C6} satisfied.
In what follows, we propose our LCG framework to implement IA. We first generate a set of beamforming vectors, then address the vector assignment problem by adopting a deep reinforcement learning (RL) approach \cite{lwd}.
\subsection{Vector Generation}
As previously described, generating the beamforming vector for one-to-one IA is straightforward. It requires a sufficient number of vectors in $\mathcal{B}(C)$, such that any $C$ of them are linearly independent. However, beamforming vectors for subspace IA usually require special design \cite{maleki2014index,suh2008interference}. Thanks to the efficiency of reinforcement learning, we do not need to design vectors in a special way as in traditional methods. Instead, we directly generate all ``0-1'' vectors of a given size $C \times 1$ excluding the all ``0'' one. This results in a total of $2^C - 1$ vectors, indicating $2^C - 1$ subspaces. 
For Example 3 in Figure \ref{4_nodes_sample_subspace_scalar}, we generate $2^3-1 = 7$ vectors: $[1 \; 0\; 0]^T, [0 \; 1\; 0]^T, [0 \; 0 \; 1]^T, [1 \; 1\; 0]^T, [1 \; 0\; 1]^T, [0 \; 1 \; 1]^T,$ and $[1 \; 1\; 1]^T$.

In fact, the total number of generated vectors does not impact the achieved DoF (given a sufficient number of vectors), as only a subset of these vectors may actually be used to satisfy the IA conditions. We denote the set of all vectors as $\mathcal{V}$, containing $S$ vectors.

\begin{figure}[h]
\begin{center}
{%
 \includegraphics[width=0.47\textwidth]{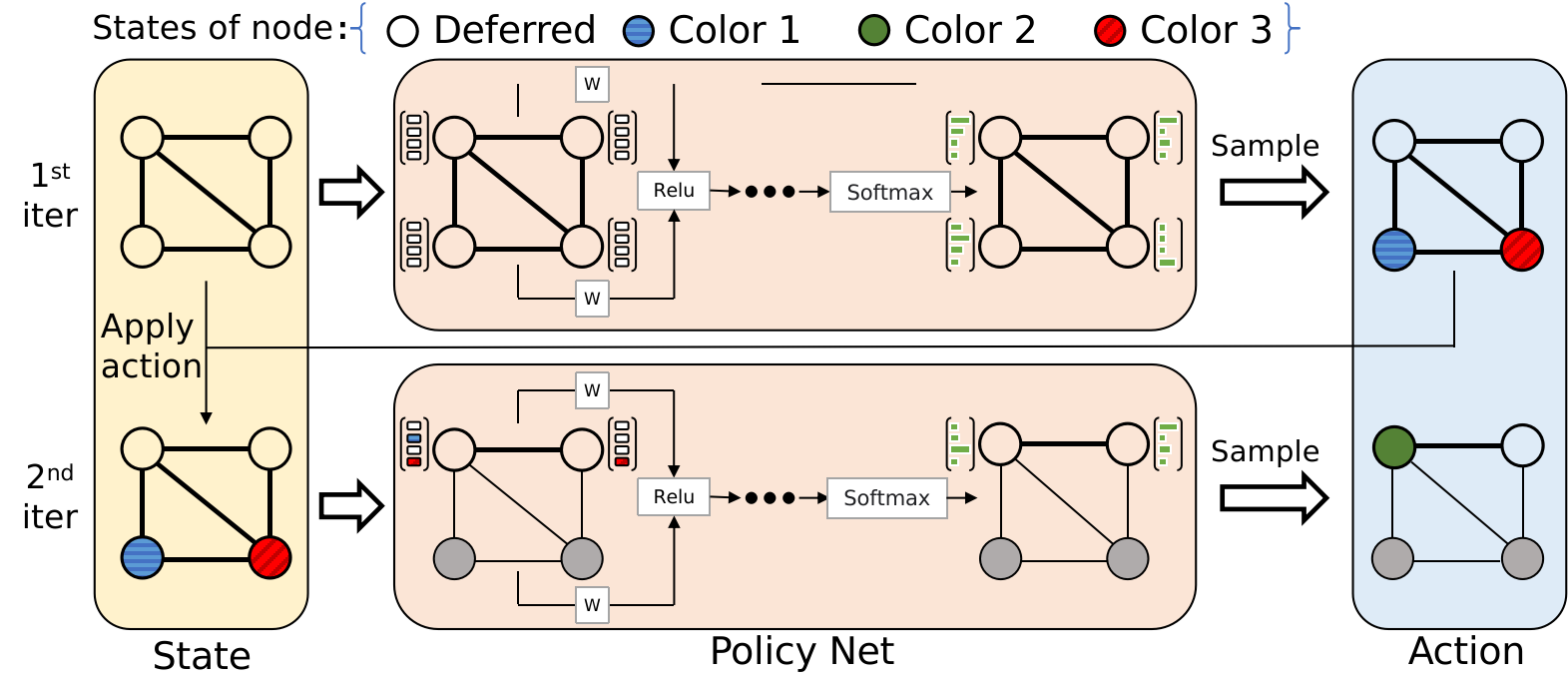}}  
\end{center}
   \caption{The iterative procedure of learning to defer for local coloring.}
   \label{iteration}
\end{figure}

\subsection{Learning to Code on Graphs}
\subsubsection{Overview of Our LCG Approach}\label{overview_LCG}
The RL approach to the vector assignment takes the directed message conflict graph $\gG_d=(\gV, \gE)$, and the number of symbol stream $b$ as the input. It attempts to produce a IA scheme that achieves the best achievable DoF value of $b/C$.

First, we utilize a \textit{C-selector} to initialize $C$. For instance, one can check the DoF outer bound (e.g. the MAIS bound \cite{bar2011index}) and choose the minimum value of $C$ such that $b/C$ is less than or equal to the outer bound. Then we repeat the RL method with increasing $C$ until a valid IA scheme is found. As shown in Figure \ref{iteration}, at each iteration, the agent (policy network) gives a vector from the vector set to some of the undetermined nodes and defers the remaining nodes to later iterations. This process will be repeated until all nodes have been given vectors. In short, the proposed LCG gradually makes decisions, and the final iteration (with early stop applied) yields the ultimate result. It's important to note that once a decision is made for a node, it remains fixed and won't be altered in subsequent iterations unless being rolled back. These decisions serve as known information to aid the model's judgment. This approach, referred to as ``learning to defer'' \cite{lwd}, allows our model to begin with simpler nodes, solidify them, and then make more challenging decisions based on the latest state.

The RL procedure can be represented as Markov decision processes (MDP) with four essential components:

\paragraph{State}
For each stage of the MDP, we define the RL state as a node-state vector $\vs = [s_j:j \in \gV]\in \{1,2,\dots,S,0\}^{\gV}$, where $s_j = c$ represents that node $j$ is assigned vector $\v_c$, $\v_c \in \mathcal{V}$, and $s_j = 0$ represents node $j$ was decided to be deferred. It is worth noting that when $b \geq 2$, each node should be assigned $b$ vectors. We employ the node splitting method to reduce dimensionality, ensuring that each node is assigned a single vector, with details presented in Appendix \ref{Node Splitting Appendix}. The node-state is initialized to all ``deferred'', i.e., $\vs_0 = [0:j \in \gV]$, and the algorithm is terminated when all nodes have been assigned vectors or have reached the iteration limit $L$.

\paragraph{Action}
Given a state $\vs$, the agent will output a corresponding action $\va_0 = [a_j:j \in \gV_0]\in \{1,2,\dots,S,0\}^{\gV_0}$ towards deferred nodes set $\gV_0$. The nodes that have been assigned in previous stages will not be given any new action. Similar to the state, the node $j$ is assigned vector $\v_c$ when $a_j = c$, or deferred when $a_j = 0$.

\begin{figure}[t]
	\begin{center}
		\includegraphics[width=0.4\textwidth]{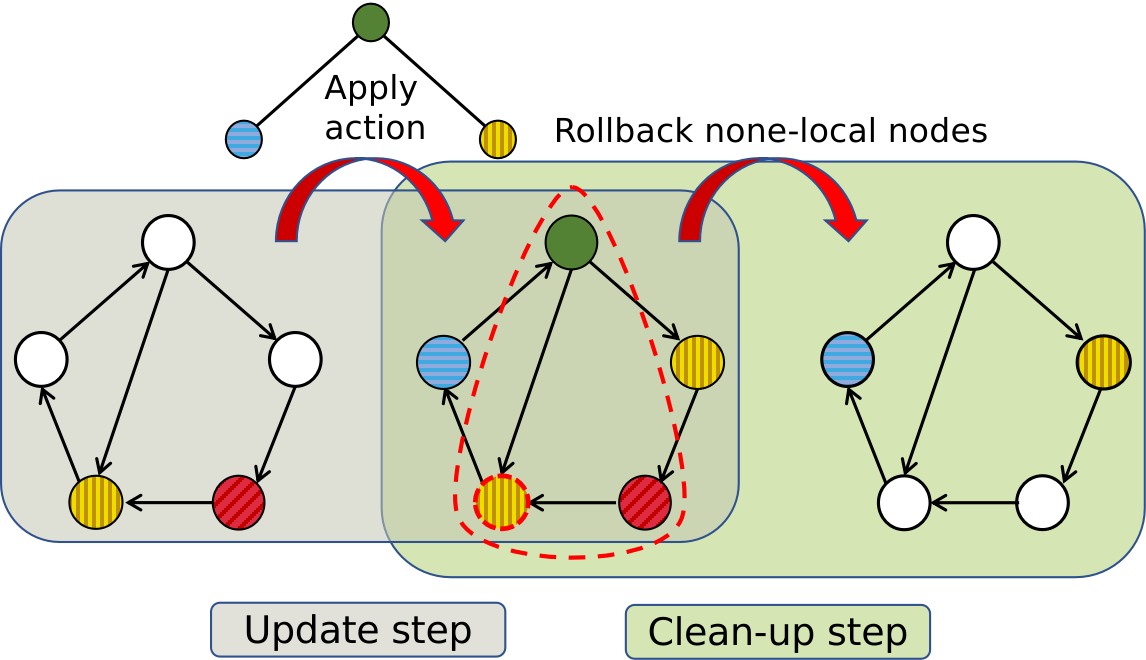}
	\end{center}
	\caption{Illustration of the transition process. Different colors represent vectors that are linearly independent. The three nodes inside the red triangle do not satisfy the IA condition Eq. \ref{IAcondition1}, so they were rolled back to the deferred stage during the clean-up phase.}
	\label{transition2}
\end{figure}

\paragraph{Transition}The RL transits from state $\vs$ to the next state $\vs'$ through two steps: \textit{update} and \cleanup.
\begin{itemize}
\item In the \textit{update} step, RL overwrites the deferred part of the previous state with the action $\va_0$, resulting in an intermediate state $\vs''$, i.e., $s''_j = a_j$ for $j \in \gV_0$ and $s''_j = s_j$ otherwise. 

\item In the \cleanup\ step, RL compute the values of $\mathrm{rank}(\S_j)$ and $\mathrm{rank}(\I_j)$ for each node $j$, then rolls back the states of the closed in-neighborhood of node $j$ to deferred if the condition Eq. \ref{IAcondition1} is not satisfied. Note that even if the node $i$ is already assigned vector in previous iterations, it remains possible to be rolled back. See Figure \ref{transition2} for a more detailed illustration of the transition between two states.
\end{itemize}

\paragraph{Reward}
We define the reward for taking an action in a state. The reward is the sum of two parts: the cardinality reward $R_{c}$, and the early-terminated reward $R_{t}$, i.e., 
\begin{IEEEeqnarray}{rCl}
R = R_{c}+ R_{t},
\end{IEEEeqnarray}
as described in Section \ref{reward}.

The policy network learns through repeated episodes (sequences of states, actions, and rewards) to adopt actions that maximize cumulative reward. Given the cumulative reward for each placement, we utilize proximal policy optimization (PPO) \cite{ppo} to update the policy network's parameters.

\subsubsection{Reward}\label{reward}
We consider two types of rewards.
\paragraph{Cardinality Reward}
Suppose the MDP transits from state $\vs$ to state $\vs'$, the cardinality reward is defined as 
\begin{IEEEeqnarray}{rCl}
R_{c}(\vs,\vs') = \sum_{i\in \gV\setminus \gV'_0} 1 - \sum_{i\in \gV \setminus \gV_0} 1.
\end{IEEEeqnarray}
This will reward the agent if more nodes are assigned in the new state. If an action causes more rollbacks than are assigned, the model gets a negative reward for this single iteration. However, if this rollback proves beneficial for future decisions, resulting in a greater total assignment count (up to the number of nodes of the graph), then such rollback would still be encouraged. By doing so, RL tends to extend the cardinality of the successfully assigned node set.

\paragraph{Early-terminated Reward}
In order to encourage the model to make decisions faster, the model is rewarded with $R_{t}$ when the algorithm terminates at time $t$ and given the iteration limit $L$, where 
\begin{IEEEeqnarray}{rCl}
R_{t} = \beta \times \frac{L-t}{L},
\end{IEEEeqnarray}
with an adjustable parameter $\beta$. Our experiments have demonstrated that this significantly increases the speed of training. 

\subsubsection{Policy and Value Network Architecture}
Our model uses Actor-Critic reinforcement learning based on graph convolutional neural networks (GCNN). Both policy network $\pi(\va|\vs)$ and value network $q(\vs,\va)$ consist of $4$-layers GraphSAGE networks \cite{graphsage} with GCN aggregator \cite{gcn}. The $n$-th layer performs
the following transformation on input $\mH$: 
\begin{IEEEeqnarray}{rCl}
h^{(n)}(\mH) = \text{ReLU}(\mH \mW_1^{(n)}+\mD^{-\frac{1}{2}} \mB \mD^{-\frac{1}{2}}\mH \mW_2^{(n)}),
\end{IEEEeqnarray}
where $\mB$ and $\mD$ represent the adjacency matrix and degree matrix, respectively. $\mW_1^{(n)}$ and $\mW_2^{(n)}$ are the weights updated during the training process. To create actions and value estimations at the final layer, the policy and value networks use softmax and graph read-out functions with sum pooling \cite{gin} instead of ReLU. The neural network's input features are the current iteration-index of the MDP and the sum of the one-hot encoding of the neighbor's state. Thanks to these features, we only take the subgraph induced on the deferred nodes $\gV_0$ as the input of the networks.

\subsubsection{Flexible Approach to Diverse IA}
We are committed to ensuring that LCG can universally produce a variety of IA schemes. Therefore, we do not train LCG for any specific IA method. Instead, we focus on training it for the fundamental aspect of IA -- vector assignment -- via graph coloring. During testing, we have the flexibility to adjust the parameters $b$ and $C$ freely and choose whether or not to use node splitting. This flexibility (also interpreted as model generalization) enables us to systematically produce various IA schemes using LCG for different network sizes and antenna configurations, at the cost of sacrificing the full potentials of the framework.
 
In order to solve graph coloring, we simply regard the vectors in the MDP as colors, and switch the \cleanup\ step to the following policy. Once RL identifies nodes that are adjacent but assigned the same color, violating the coloring rules, these nodes are then mapped back to ``deferred''. During training, we choose the number of total colors $S$ as the chromatic number $\chi(\gG)$ of input graphs, which represents the smallest number of colors needed to color the graph. For testing, one can utilize the model trained with any $S\geq \chi(\gG)$ to ensure that the number of total vectors is sufficient.

\begin{figure*}
   \centering
   \subfloat[Ex.~\ref{SVIA_example_section}: An instance with \textit{directed triangle} and a SVIA solution.]{%
     \label{SVIA_example}
     \includegraphics[width=4.4cm]{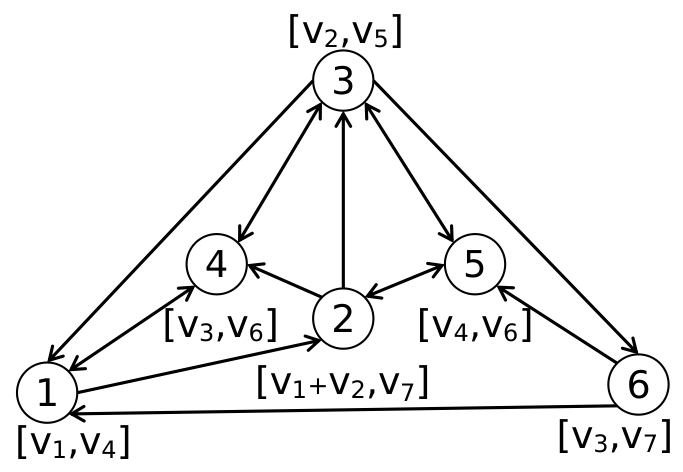}}  
     \hfil
   \subfloat[Ex. \ref{odd_hole_exe}: An instance with \textit{odd hole} and a SSIA solution.]{%
     \label{Ex. 11: SVIA_oddhole_2}
     \includegraphics[width=4.2cm]{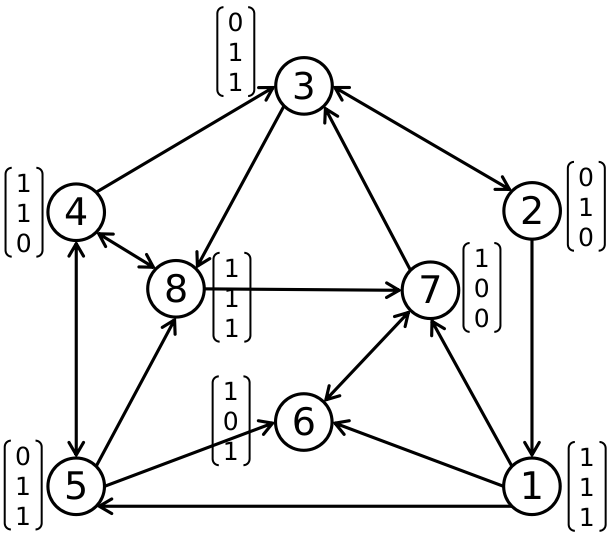}}  
        \hfil
   \subfloat[Ex. \ref{4clique_exe}: An instance with \textit{4-clique} and a SSIA solution.]{%
     \label{Ex. 12: SVIA_4clique_1}
     \includegraphics[width=4.3cm]{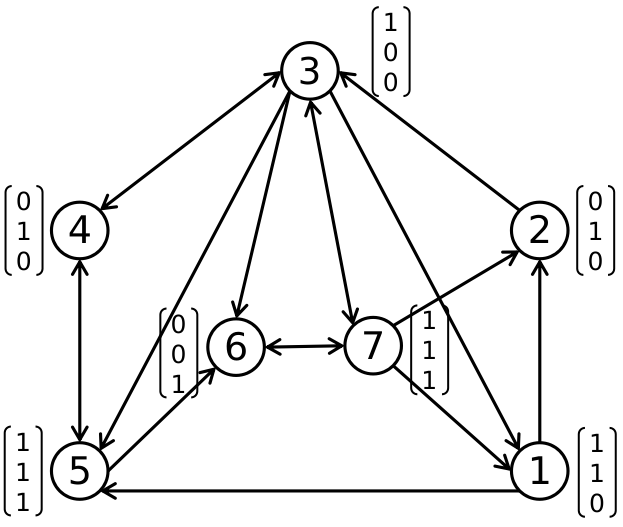}}  
        \hfil
   \caption{(a) An instance containing \textit{directed triangle} structure and a SVIA solution achieving DoF 2/7, (b) an instance containing \textit{odd hole} structure achieving optimal DoF 1/3, and (c) an instance containing \textit{4-clique} structure achieving optimal DoF 1/3.}
   \label{discussion_structure}
   \vspace{-20pt}
\end{figure*}

\section{Discussions}\label{Discussions}
In this section, we present several intriguing TIM instances discovered by LCG, whose topological structures may inspire future endeavors in this line of research.
In our exploration using LCG, a multitude of instances have been identified where subspace IA surpasses one-to-one IA in achieving higher DoF. We try to explore some potential regularities that could serve as inspiration for the design of IA solutions in future research endeavors. The following three discovered structures are highlighted. 

\paragraph{Directed triangle}\label{Directed triangle Discussion}The first intriguing topological structure is a $4$-clique, containing a directed triangle, where all three nodes of the triangle point towards the fourth node. This is already shown in Figure \ref{Ex. 5 SSIA}, where subspace IA is required to achieve the optimal DoF of 1/3.
\begin{exe}\label{SVIA_example_section}
Figure \ref{SVIA_example} is another instance containing the \textit{directed triangle} structure, but there is still some gap between subspace IA and the MAIS outer bound. In particular, nodes 1, 2, and 3 form a directed triangle, as there exist edges (1,2), (2,3), and (3,1), and all point to node 4. So it requires the subspace IA method to handle the $4$-clique 1-2-3-4. Specifically, we assign vectors as shown in the figure, with $\v_1, \dots, \v_7$ being $7$-dimensional linearly independent vectors, achieving the DoF value of $2/7$. 
However, both of the MAIS and MICD bounds indicate that the symmetric DoF is upper bounded by $1/3$, i.e., $2/7 \le d_{\mathrm{sym}} \le 1/3$. 
While the LCG framework is not able to produce better IA solution with higher DoF, it is unclear if the outer bound is tight.
\hfill $\square$
\end{exe}

\paragraph{Odd Hole} 
The second structure is odd hole, defined as a hole with an odd number (at least five) of nodes.
A hole is an undirected chordless cycle, which is a cycle in an undirected graph such that no two nodes of the cycle are connected by an edge (i.e., chord) that is not part of the cycle itself. 
\begin{exe} \label{odd_hole_exe}
The example presented in Figure \ref{Ex. 11: SVIA_oddhole_2} contains odd holes 1-2-3-4-5 and 1-7-3-4-5. The detailed vector generation and assignment are automatically produced by LCG and shown in the figures. The SSIA solution achieves the optimal DoF of 1/3, according to the MAIS bound. \hfill $\square$
\end{exe}

\paragraph{4-Clique}
The third structure is a $4$-clique, say $u$-$x$-$y$-$z$, where $u$ is pointed by two nodes in this clique, say $x, y$. There exists another node $v$ point to $u$, and $u$, $v$, $z$ all point to the same new node.
\begin{exe}\label{4clique_exe}
The example presented in Figure \ref{Ex. 12: SVIA_4clique_1} contains a $4$-clique 1-2-3-7, where 2 and 7 point to 3. Another node, 4, point to 3 as well, and nodes 3, 4, and 7 all point to 5. The SSIA solution with detailed vector generation and assignment are automatically executed by LCG and illustrated in the figures. According to the MAIS bound, this SSIA solution achieves the optimal DoF of 1/3.
\hfill $\square$
\end{exe}

The above three examples suggest that: 
\begin{itemize}
    \item [1)]
    For some special instances, more advanced subspace vector IA coding schemes and/or tighter outer bounds than MAIS/MICD may be required for optimality.
    \item [2)]
    There might exist a family of TIM instances with some common topological structures such that subspace scalar IA coding schemes can be proven DoF optimal.
\end{itemize}

\section{Experiments}
In this section, we present the results obtained by applying the proposed LCG to various datasets, particularly exploring the maximum DoF achievable for each instance and the required IA techniques. Additionally, we investigate the improvements brought by SIMO networks compared to SISO networks. Finally, we evaluate the effectiveness and generalization capability of LCG.\footnote{The detailed experiment setups and  implementation are available at \url{https://github.com/ZhiweiShan/Learning-to-Code-on-Graphs}}

\textbf{Dataset generation}:
{
We generate two types of random graphs. One is Erdős-Rényi (ER) random graphs \cite{batagelj2005efficient} as message conflict graphs of TIM instances, and the other one is device-to-device network topologies with random devices locations (Wireless Net) \cite{yi2015itlinq+} following a short-range outdoor channel model ITU-1411 \cite{ITU-R-P1411-8}. ER graphs are widely used as random graphs, including a wide range of representative graph structures. The Wireless Net graphs are transformed into message conflict graphs, as shown in \ref{TIM Representations}.}

\textbf{LCG with RL}:
In our LCG approach, we generate a total of 50,000 graphs for training employing various random parameter settings. These graphs are subsequently categorized into multiple datasets based on their chromatic numbers. We use the datasets with chromatic number equals to $S$ to train LCG with number of colors $S$. The hyperparameters utilized in our approach can be found in Appendix \ref{Implementation_of_L2C}.

\begin{figure}[htbp]
	\begin{center}
 \includegraphics[width=0.45\textwidth]{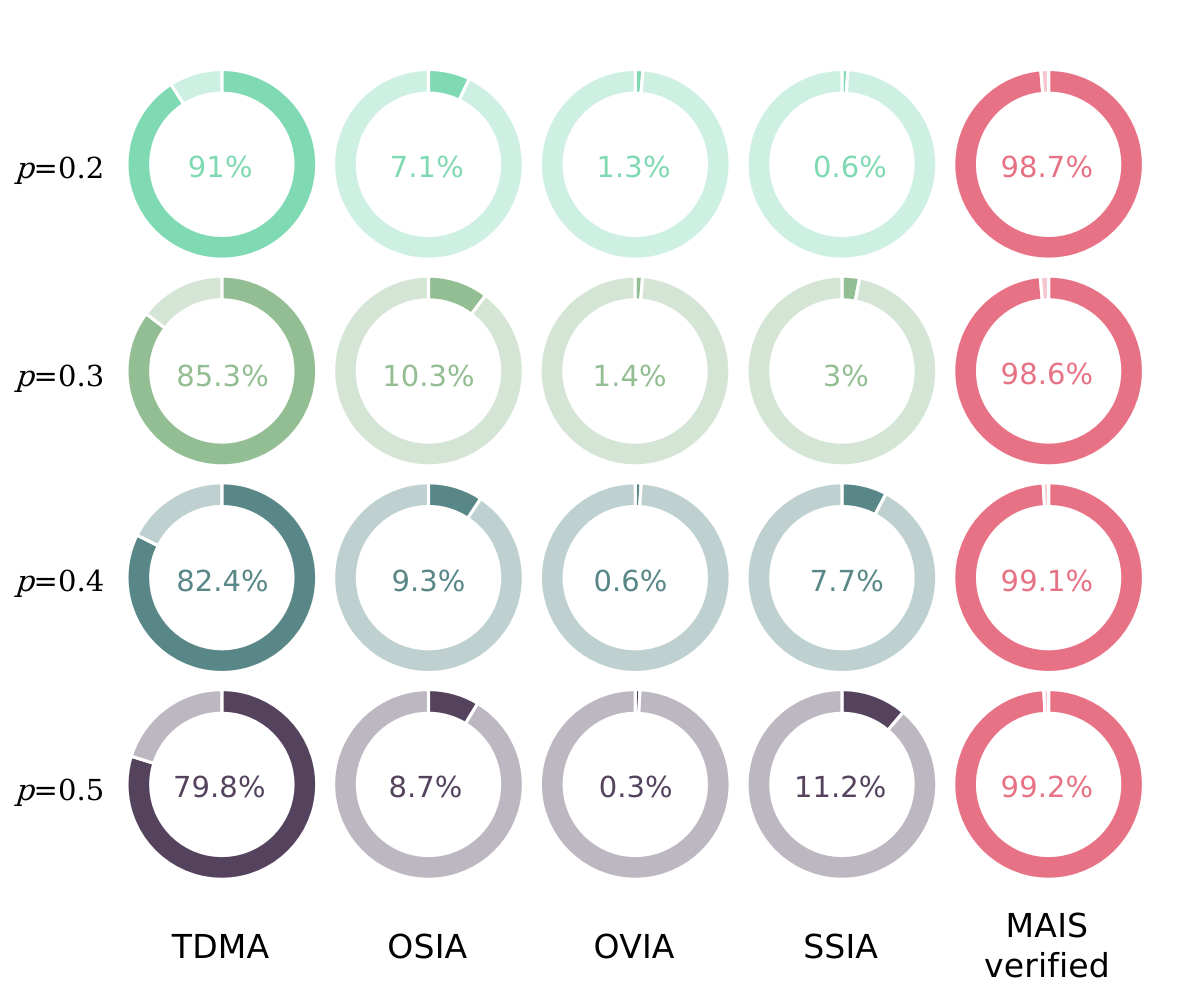}
	\end{center}
	\caption{The proportion of different methods that yield the best achievability. Each row presents outcomes for ER random graphs with $p=0.2, 0.3, 0.4$, and $0.5$. The percentages in the first four columns denote the proportion of instances that the specified method (as the simpliest one) achieves the maximal DoF. The final column indicates the proportion of instances achieving DoF agree with the MAIS outer bound, thereby confirming optimality.}
	 \label{ER_IA_proportion}
\end{figure}
\begin{figure}[htbp]
	\begin{center}
 \includegraphics[width=0.45\textwidth]{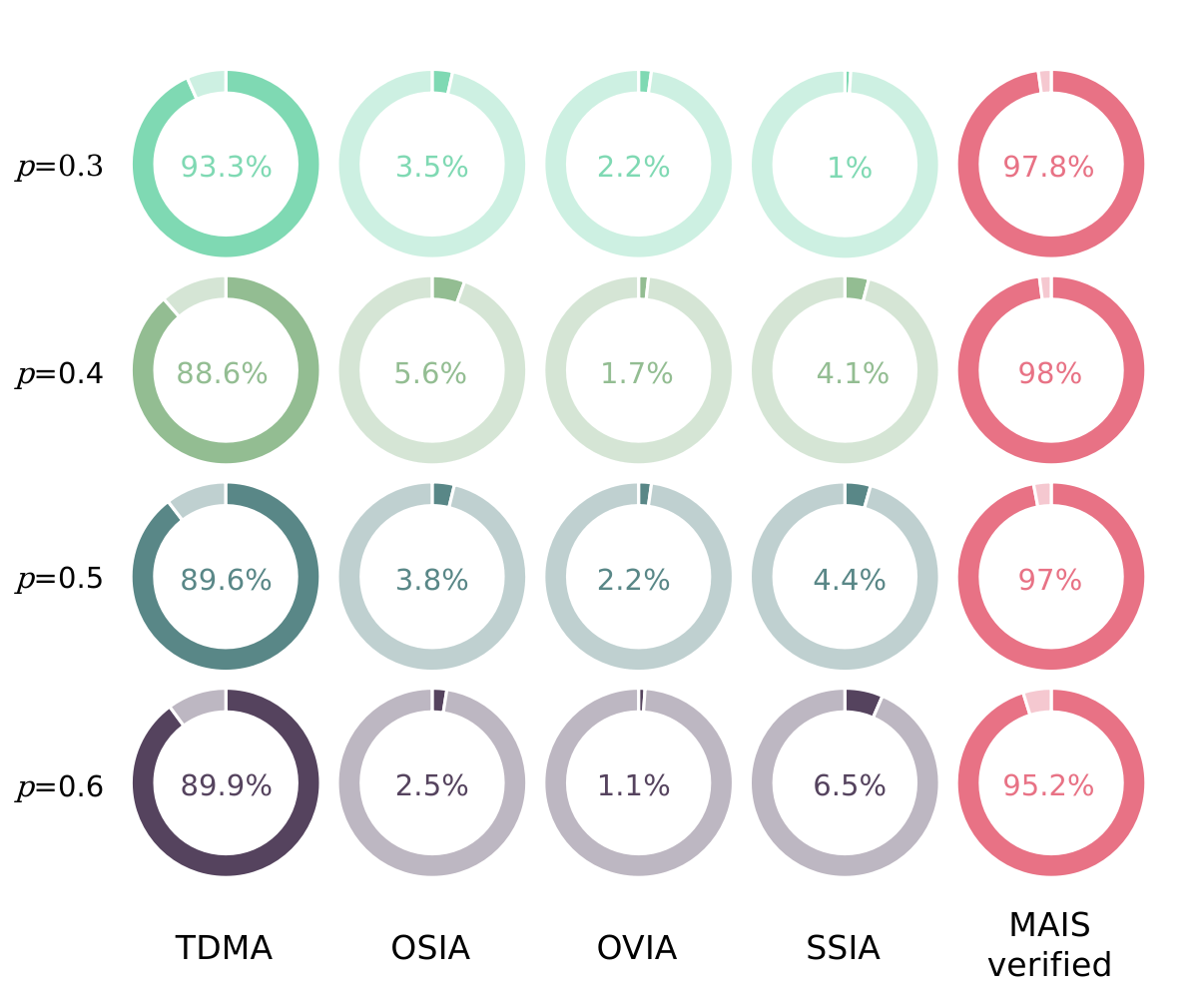}
	\end{center}
	\caption{The proportion of different methods that yield the best achievability. Each row presents outcomes for Network graphs with $p=0.3, 0.4, 0.5$, and $0.6$. The percentages in the first four columns denote the proportion of instances that the specified method (as the simpliest one) achieves the maximal DoF. The final column indicates the proportion of instances achieving DoF agree with the MAIS outer bound, thereby confirming optimality.}
	\label{Network_IA_proportion}
\end{figure}

\subsection{Evaluation of Different Coding Schemes Produced by LCG}
\subsubsection{Achievability of Various IA Schemes in SISO Setting}
We first consider the four linear coding schemes, i.e., orthogonal access (e.g., TDMA), OSIA, OVIA, and SSIA, produced by the proposed LCG framework, and evaluate their achievability on four datasets of ER random graphs. Each dataset contains 1000 directed graphs with 6 nodes and edge probabilities of $p=0.2$, $0.3$, $0.4$, and $0.5$, respectively. We documented the method employed by each TIM instance to achieve the maximal DoF among the four methods. If for an instance the same DoF value can be achieved by multiple methods, we attribute it to the simplest method, according to relations in Eqs.~(\ref{eq:IA-relation1}) and (\ref{eq:IA-relation2}). Figure \ref{ER_IA_proportion} demonstrates the proportion of TIM instances that the specific method achieves the highest DoF value in the first four columns. In the last column, we display the proportion of TIM instances where the achieved DoF value agrees with the MAIS outer bound, i.e., the optimal DoF is achievable. The optimality gap that the MAIS bound can not verify may be attributed to the loose outer bound and/or the need of advanced coding techniques.

It can be seen that most instances (80\%-90\%) in the four datasets can achieve the best DoF using TDMA. Quite a few (7\%-10\%) instances require the OSIA method, while only a small percentage of instances need the OVIA method to achieve the best DoF. Note that as the graph density increases from 0.2 to 0.5, the proportion of instances that require the SSIA method to achieve the best DoF increases from 0.6\% to 11.2\%, suggesting that the SSIA method contributes more to dense graphs. In addition, with the help of these methods, the majority of instances (about 99\%) have reached the optimum, while there is only approximate 1\% of instances whose optimality can not be verified by the MAIS outer bound.

We then conducted the same evaluation on the D2D wireless network. We generated four datasets, each containing 1000 graphs with 8 nodes and densities (of TIM topology) $p=0.3$, $0.4$, $0.5$, and $0.6$ respectively. As shown in Figure \ref{Network_IA_proportion}, on these datasets, the proportion of instances reaching the best DoF with TDMA is higher, approximately 90\%, while the proportion requiring advanced methods decreases accordingly. However, the proportion of instances that unable to reach the MAIS bound increases with increasing density.

\begin{figure}[htbp]
	\begin{center}
 \includegraphics[width=0.5\textwidth]{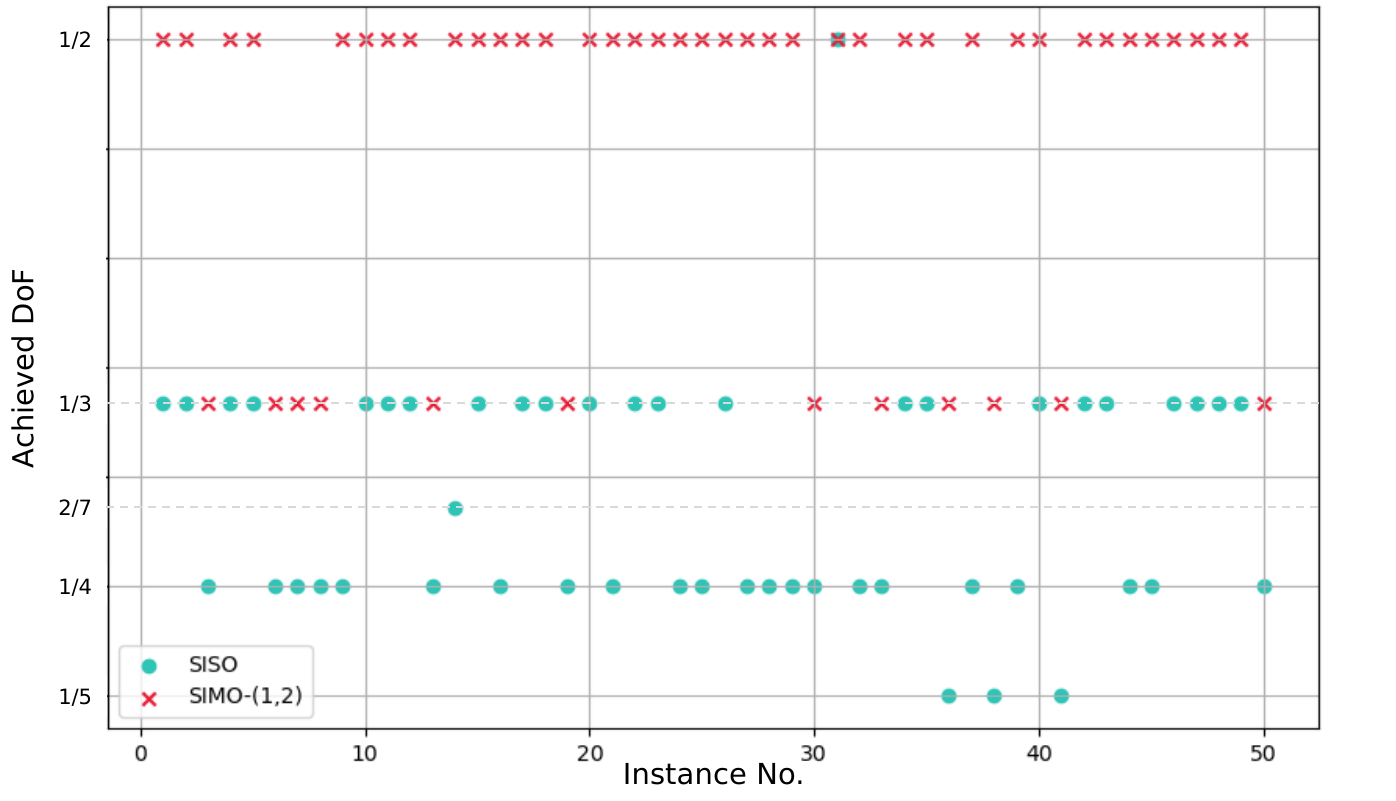}
	\end{center}
	\caption{Achieved DoF under SISO and SIMO-(1,2) network settings.}
	\label{SISO_SIMO_compare}
\end{figure}

\subsubsection{Comparison between SISO and SIMO Settings}
In this section, we compare the achievable DoF between SISO and SIMO networks. Specifically, we generated 1000 ER random graphs with 6 nodes and $p=0.4$. Then we ran LCG under both SISO and SIMO-$(1,2)$ settings. We tried various IA methods under the SISO setting, consistent with the previous section, and recorded the best result as the achieved DoF. For the SIMO setting, we employed scalar IA. The best achievable DoF for both settings are depicted in Figure \ref{SISO_SIMO_compare} (only partially). 
The majority (98\%) of instances benefit from the additional receive antennas, resulting in substantial improvements in achievable DoF values. About 35\% of instances even exhibited a twofold increase in their DoF.

\subsection{Evaluation of the LCG Framework}
As our proposed LCG is the first learning-to-code algorithm for topological IA to the best of the authors' knowledge, there are no appropriate baselines available for comparison. Therefore, given that graph coloring serves as the backbone of LCG, 
{we aim to verify whether LCG can achieve the optimal coloring, i.e., using the minimum number of colors to color the graph, while also evaluating its efficiency by comparing it with traditional graph coloring algorithms.}

{We classify the ER and Wireless Net graphs through calculate their chromatic numbers $\chi(\gG)$ by solving linear programming problems using Gurobi \cite{gurobi}. The size represents the number of nodes in the message conflict graph. For a comprehensive understanding of the graph instance generation process, we refer the reader to Appendix \ref{Dataset_Details}.}

\subsubsection{Performance and Running Time Comparison}
To assess the quality of the solutions, we measure the optimal ratio, which represents the proportion of graphs that can be colored using the chromatic numbers of colors $\chi(\gG)$ by each method.

\textbf{Baselines}:
We compare our method to two heuristic algorithms: the smallest-last greedy method with the interchange (SLI) \cite{matula1983smallest, deo2006interchange} and TabuCol \cite{hertz1987using}. The smallest-last greedy method is a simple yet powerful approach. It assigns each node in a sequential manner to the lowest indexed color that does not result in any conflicts, and it adds new colors when necessary. The interchange technique is employed to enhance the effectiveness of any sequential coloring algorithm.
TabuCol, on the other hand, is a well-studied heuristic based on Tabu local search. In this method, the chromatic number is assumed to be given, and we set the maximum number of iterations for TabuCol to 1000.

The results are presented in Table \ref{performance}. Notably, the proposed LCG model consistently achieves the highest optimal ratios across all datasets. Moreover, LCG exhibits superior computational efficiency compared to the SLI greedy method in most cases. All our experiments are performed using a single GPU (NVIDIA A100) and a single CPU (AMD EPYC 7452).

\begin{table}[htbp]
	\centering
	\caption{Optimal ratio on test graphs, where the best ratios are marked in bold. Running times (in seconds) are provided in brackets. In this table, training and testing use data from the same distribution. }
	\label{performance}
	\begin{footnotesize}		
 \resizebox{0.98\columnwidth}{!}{
		\begin{tabular}{ccc|ccc}
			\hline
			Type                          & size                      & $\chi$ & SLI         & TabuCol     & LCG         \\ \hline
			\multirow{4}{*}{ER}           & \multirow{2}{*}{15} & 5      & 0.98 (1.83) & \textbf{1} (9.58)    & \textbf{1} (2.38)    \\
			&                        & 6      & 0.99 (0.68) & \textbf{1} (2.53)    & \textbf{1} (1.79)    \\ \cline{2-6} 
			& \multirow{2}{*}{30} & 7      & 0.73 (7.28) & 0.87 (1677) & \textbf{0.92} (5.10) \\
			&                        & 8      & 0.85 (15.8) & 0.92 (2699) & \textbf{0.94} (8.59) \\ \hline
			\multirow{4}{*}{Wireless Net} & \multirow{2}{*}{15} & 5      & 0.94 (9.64) & \textbf{1} (406)     & \textbf{1} (5.06)    \\
			&                        & 6      & \textbf{1} (10.6)    & \textbf{1} (169)     & \textbf{1} (5.41)    \\ \cline{2-6} 
			& \multirow{2}{*}{30} & 7      & 0.88 (28.5) & 0.97 (3602) & \textbf{0.99} (13.2) \\
			&                        & 8      & 0.99 (10.8) & 0.99 (497)  & \textbf{1} (6.36)    \\ \hline
		\end{tabular}}
	\end{footnotesize}
\end{table}

\subsubsection{Generalization and Transferability}
{
Finally, we evaluate the generalization ability of our method by examining its performance on ER graphs of varying sizes. Specifically, we train the LCG model on ER graphs of certain sizes and test its generalization ability exclusively on ER graphs of different sizes. The comprehensive results are presented in Table \ref{Generalization}.
}

{
Each column represents the optimal ratio of models trained on different training datasets and tested on a specific dataset. It can be observed that the results in each column are similar, indicating that the model is not overly sensitive to the training dataset. We note that when the test size is 30, the model trained on data of the same size performs worse than the others. This may be due to the fact that the training dataset with size 30 is denser, which makes the training process more challenging for the neural network. Utilizing techniques such as edge dropout or attention mechanisms could potentially improve performance in this case.
}

\begin{table}[htbp]
	\centering
	\caption{Optimal ratio on graphs from different types and size graphs. Each dataset is filtered from 5000 test graphs by $\chi=7$.}
	\label{Generalization}
	\resizebox{0.8\columnwidth}{!}{
	\begin{tabular}{l|ccccccc}
		\hline
		\diagbox[width=2cm,height=1cm]{Train\\ \ size}{Test\\size} &
		\begin{tabular}[c]{@{}c@{}}15\end{tabular} &
		\begin{tabular}[c]{@{}c@{}}20\end{tabular} &
		\begin{tabular}[c]{@{}c@{}}25\end{tabular} &
		\begin{tabular}[c]{@{}c@{}}30\end{tabular} &\\ \hline
		\qquad15 & 0.999 & 0.995 & 0.996 & 0.934  \\ \hline
		\qquad20 & 1     & 0.993 & 0.996 & 0.944  \\ \hline
		\qquad25 & 0.997 & 0.993 & 0.996 & 0.928  \\ \hline
		\qquad30 & 1     & 0.993 & 0.995 & 0.925 \\ \hline
	\end{tabular}
}
\end{table}

\section{Conclusion}
\label{sec:conclusion}
Revisiting topological interference management (IA) from a vector assignment perspective, we proposed a learning-to-code on graphs (LCG) framework for the topological interference management (TIM) problem, by leveraging graph reinforcement learning to assign vectors on the message conflict graph with certain decision-making policies. 
The proposed LCG framework can not only recover the known one-to-one IA solutions to a wide range of network topologies, but also discover new yet hard-to-craft subspace IA solutions for some special cases.
A comprehensive experimental evaluation of the proposed framework demonstrates its efficacy and efficiency in learning IA solutions in a systematic way.
As the first learning-to-code approach to the TIM problem, there are still great potentials to develop new models and algorithms in the future work to discover advanced coding schemes for large-scale network topologies and the multiple-antenna cases.

\appendices
\section{Node Splitting}\label{Node Splitting Appendix}
For vector IA, we utilize the technique of node splitting to transform vector IA problem in the original graph into a more conventional scalar IA problem on a reconstructed graph. Each node undergoes careful splitting, with a vector assigned to every node within the splitting graph. Then, we merge the split nodes back into their original nodes, obtaining a vector IA scheme for the original graph. We then define the $b$-order node splitting graph.\footnote{The $b$-order node splitting graph can be seen as the Cartesian product of the original conflict graph and the directed clique of size $b$.}

\begin{defn}[$b$-order node splitting graph]
Let $\gG = (\gV, \gE)$ be a directed graph, and let $b$ be a positive integer. The $b$-order node splitting graph of  $\gG$, denoted as $\gG_b' = (\gV_b', \gE_b')$, is defined as follows:
Each original node $v \in \gV$ is split into $b$ nodes $v_1, v_2, \ldots, v_b$ such that 
\begin{align}
    \gV_b' = \{v_1, v_2, \ldots, v_b \mid v \in \gV \}.
\end{align}
For each edge $(u, v) \in \gE$, there exist $b^2$ edges $\{(u_i, v_j) \mid 1 \leq i,j \leq b \}$ in $\gE_b'$. Moreover, the $b$ split nodes of each original node are fully connected. That is, 
\begin{align}
    \gE_b' = &\{(u_i, v_j) \mid (u, v) \in \gE, 1 \leq i,j \leq b \} \\
    &\cup \{(v_i, v_j) \mid v \in \gV, 1 \leq i,j \leq b, i \neq j\}.
\end{align}
\end{defn}

In other words, the $b$-order node splitting graph is obtained by splitting each node in the original graph into $b$ directed fully connected nodes, and then connecting each split node of $u$ to each split node of $v$, if $u$ points to $v$ in the original graph. 

\begin{figure}[htbp]
	\begin{center}
		\includegraphics[width=0.47\textwidth]{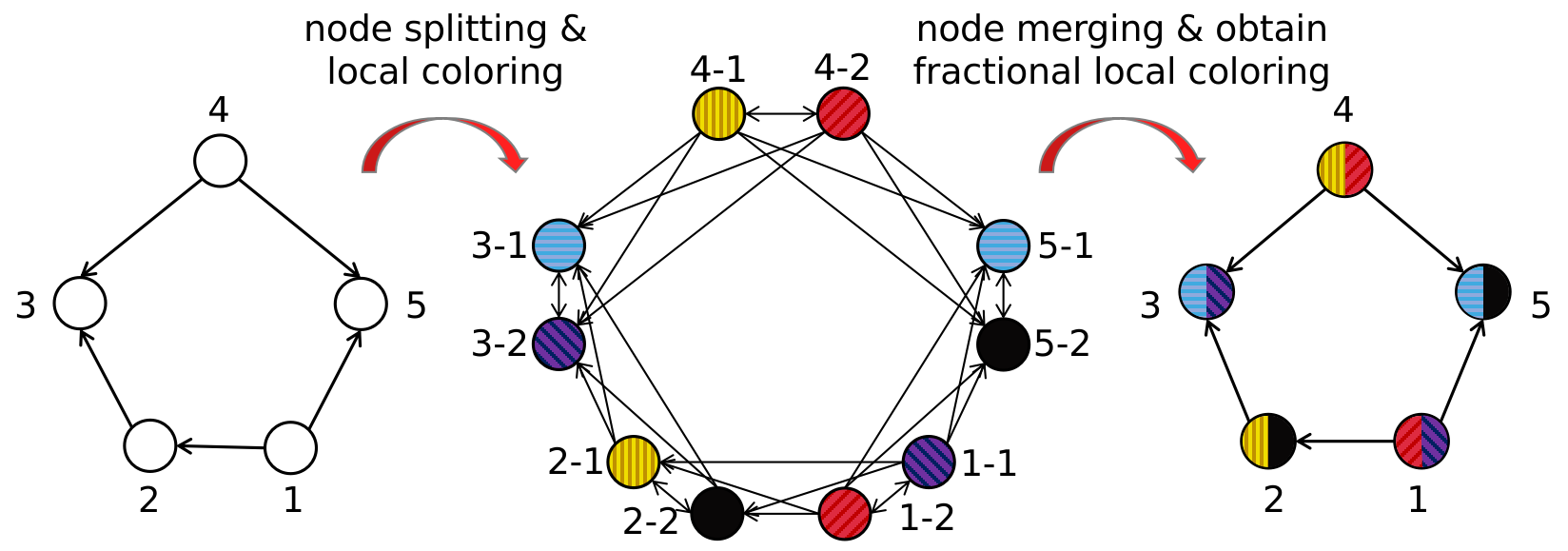}
	\end{center}
	\caption{Illustration of node splitting and merging.}
	\label{node_splitting_picture}
\end{figure}

Figure \ref{node_splitting_picture} showcases an example for the generation of a $2$-order node splitting graph, the subsequent application of local coloring on the splitting graph, and the subsequent merging of the splitting graph and coloring scheme back into the original graph to yield a fractional local coloring scheme. 

In essence, we have expanded the scope of the LCG method to address the vector IA problem by employing a combination of node splitting and merging techniques.

\section{Dataset Details}\label{Dataset_Details}
We train on graphs randomly generated based on variant specific parameters. 50,000 graphs were generated and separated according to the chromatic number for training, and 5,000 graphs for evaluation.
\begin{itemize}
    \item  The specific parameters for random graphs in Table \ref{performance},
    \ref{Generalization} are shown in Table \ref{Table_1_parameter}, \ref{Table_2_parameter} separately. $q$ represents the percent of randomly choosing demanded messages.
    
    \item For Wireless Net, we randomly distribute transmitters and receivers within a square area of 1,000 m $\times$ 1,000 m.  As in \cite{yi2015itlinq+}, the simulated channel follows the LoS model in ITU-1411. The carrier frequency is 2.4 GHz, antenna height is 1.5 m, and the antenna gain per device is -2.5 dB. The noise power spectral density is -174 dBm/Hz, and the noise figure is 7 dB. Each pair of transmitter and receiver is uniformly spaced within 2-65 meters. Each link is expected to operate over a 10 MHz spectrum and the maximum transmit power is 30 dBm. 
\end{itemize}
 
\begin{table}[htbp]
\centering
\caption{Specific parameters for generating graph in Table \ref{performance}. Wireless Net follows all-unicast setting, therefore q does not apply.}
\label{Table_1_parameter}
\resizebox{0.9\columnwidth}{!}{
\renewcommand{\arraystretch}{0.8}
\begin{tabular}{cc|c|cll}
\hline
Type & N                         & q & \multicolumn{3}{c}{Specific parameters}                                           \\ \hline
\multirow{4}{*}{ER} &
  \multirow{2}{*}{15} &
  \multirow{4}{*}{0.2} &
  \multicolumn{3}{c}{\multirow{2}{*}{Probability for edge creation = 0.2}} \\
     &                           &   & \multicolumn{3}{c}{}                                                              \\ \cline{2-2} \cline{4-6} 
     & \multirow{2}{*}{30} &   & \multicolumn{3}{c}{\multirow{2}{*}{Probability for edge creation = 0.2}}          \\
     &                           &   & \multicolumn{3}{c}{}                                                              \\ \hline

\multirow{4}{*}{\multirow{2}{*}{\begin{tabular}[c]{@{}c@{}}Wireless\\ Net\end{tabular}}} &
  \multirow{2}{*}{15} &
  \multirow{4}{*}{\textbackslash{}} &
  \multicolumn{3}{c}{\multirow{2}{*}{\begin{tabular}[c]{@{}c@{}}Topological density = 0.4\end{tabular}}} \\
     &                           &   & \multicolumn{3}{c}{}                                                              \\ \cline{2-2} \cline{4-6} 
 &
  \multirow{2}{*}{30} &
   &
  \multicolumn{3}{c}{\multirow{2}{*}{\begin{tabular}[c]{@{}c@{}} Topological density = 0.3\end{tabular}}} \\
     &                           &   & \multicolumn{3}{c}{}                                                              \\ \hline
\end{tabular}
}
\end{table}

\begin{table}[htbp]
\centering
\caption{Specific parameters for generating graph in Table \ref{Generalization}.}
\label{Table_2_parameter}
\resizebox{0.9\columnwidth}{!}{
\begin{tabular}{cc|c|cll}
\hline
Type & N        & q & \multicolumn{3}{c}{Specific parameters}                          \\ \hline
\multirow{4}{*}{ER} &
  15 &
  \multirow{4}{*}{0.2} &
  \multicolumn{3}{c}{Probability for edge creation = 0.3} \\ \cline{2-2} \cline{4-6} 
     & 20 &   & \multicolumn{3}{c}{Probability for edge creation = 0.25}         \\ \cline{2-2} \cline{4-6} 
     & 25 &   & \multicolumn{3}{c}{Probability for edge creation = 0.2}          \\ \cline{2-2} \cline{4-6} 
     & 30 &   & \multicolumn{3}{c}{Probability for edge creation = 0.2}          \\ \hline
\end{tabular}
}
\end{table}

\section{Implementation of LCG}\label{Implementation_of_L2C}
We use the same hyperparameters for each experiments. We set the maximum number of iterations per episode $L = 32$. Policy and value networks were parameterized using a graph convolutional network with four layers and 128 hidden dimensions. Using the Adam optimizer with a learning rate of 0.001, each instance of the model was trained for 5,000 iterations of proximal policy optimization \cite{ppo}. For each instance, we calculate 20 cases in parallel and select the best results to report. The gradient norms were clipped by $0.2$. As a function of the number of nodes in a dataset, the cardinality reward is normalized by this number. 

\IEEEtriggeratref{100}

\bibliographystyle{IEEEtran}
\bibliography{paper} 

\end{document}